\documentclass[11pt]{article} 
\usepackage[utf8]{inputenc} 

\usepackage{geometry} 
\usepackage{graphicx} 
\usepackage[parfill]{parskip} 
\usepackage{color}
\usepackage{booktabs} 
\usepackage{array} 
\usepackage{paralist} 
\usepackage{verbatim} 
\usepackage{subfig} 
\usepackage{authblk}

\usepackage{fancyhdr} 
\usepackage{sectsty}
\allsectionsfont{\sffamily\mdseries\upshape} 

\usepackage[nottoc,notlof,notlot]{tocbibind} 
\usepackage[titles,subfigure]{tocloft} 

\newcommand{\be}{\begin{equation}}
\newcommand{\ee}{\end{equation}}
\newcommand{\bea}{\begin{eqnarray}}
\newcommand{\eea}{\end{eqnarray}}

\newcommand{\met}{\slash \hspace{-0.1in} E_T }

\title{Dark Matter on Top}
\author[1]{M.A. G\'omez\thanks{miguel.gomezramirez@mavs.uta.edu}}
\author[1]{C.B. Jackson \thanks{chris@uta.edu}}
\author[2]{G. Shaughnessy \thanks{gshau@hep.wisc.edu}}
\affil[1]{Dept. of Physics, University of Texas at Arlington, Arlington, TX 76019, USA}
\affil[2]{Dept. of Physics, University of Wisconsin, Madison, WI 53706, USA}

\date{} 

\begin{document}

\maketitle

\begin{abstract}
We consider a simplified model of fermionic dark matter which couples exclusively to the right-handed top quark via a renormalizable interaction with a color-charged scalar.  We first compute the relic abundance of this type of dark matter and investigate constraints placed on the model parameter space by the latest direct detection data.  We also perform a detailed analysis for the production of dark matter at the LHC for this model.  We find several kinematic variables that allow for a clean signal extraction and we show that the parameter space of this model will be well probed during LHC Run-II.  Finally, we investigate the possibility of detecting this type of dark matter via its annihilations into gamma rays.  We compute the continuum and the line emission (which includes a possible ``Higgs in Space!'' line) and its possible discovery by future gamma-ray telescopes.  We find that the annihilation spectrum has distinctive features which may distinguish it from other models. 
\end{abstract}

\newpage

\section{Introduction}

According to the latest cosmological measurements, approximately eighty percent of the matter in the Universe is in the form of a mysterious substance called {\it dark matter}.  Despite this large abundance, we have yet to uncover the particle (or particles) which make up dark matter.  This has not been for a lack of trying, however, as over the past twenty years, dozens of experiments have searched for signals of particle dark matter.  None of which have found an {\it undeniable} signature of particle dark matter.  So far, we have only been able to state what dark matter is {\it not}:  it is definitely {\it not} a particle from the Standard Model (SM) and it is beginning to look like it is {\it not} a particle from some of the most popular extensions of the SM.

Instead of working with complicated UV-complete models, recently, many theorists have focused on models which follow a more {\it phenomenlogical approach}.  In these types of models, one only adds to the SM the minimal content needed to account for dark matter \cite{Beltran:2008xg,Fan:2010gt,Fitzpatrick:2012ix,Beltran:2010ww,Goodman:2010yf,Bai:2010hh,
Goodman:2010ku}.  The strength of these ``effective'' (or ``simplified'') models is that they encompass the interactions and parameter spaces of well-motivated models such as supersymmetry or extra-dimensional models, as well as new scenarios (which are not realized in more UV-complete models).  The typical approach in these scenarios is to introduce a single particle responsible for the current abundance of dark matter and couple it to the SM via interactions of the form:
\bea
\Delta {\cal L} \sim \frac{1}{\Lambda^n} {\cal O}_{SM} {\cal O}_{DM} \,,
\label{eq:EffDM}
\eea
where ${\cal O}_{SM}$ (${\cal O}_{DM}$) are operators constructed from SM (DM) fields and $\Lambda$ is the cut-off of the effective theory.  To ensure stability, it is usually assumed that there is a discrete symmetry present which keeps the DM particle from decaying.  Thus, the operator ${\cal O}_{DM}$ must consist of even powers of the DM field.  The most widely studied interactions of this type are the non-renormalizable interactions of the form:
\bea
\Delta {\cal L} \sim \frac{1}{\Lambda^n} \left| {\rm SM} \right|^2  \left| {\rm DM} \right|^2 \,.
\label{eq:EffDM2on2}
\eea
The nice feature of models of this form is that they result in complimentary probes of the available parameter space from direct detection and collider experiments because the same operator is responsible for the signals at these experiments \cite{Goodman:2010yf,Bai:2010hh,Goodman:2010ku,Birkedal:2004xn,Feng:2005gj,Cao:2009uw,
Cheung:2012gi,Dreiner:2013vla}.  The downside to these models is that collider bounds of higher-dimensional operators probe scales $\Lambda$ which are of the order of the collisional energies.  Thus, the higher-scale physics (i.e., the UV completion) becomes important for collider phenomenology.  This negates the whole idea behind ``simplified'' models of DM.

The way around this problem is to only consider {\it renormalizable} interactions between the SM and DM.  In fact, interactions of this type are a major component of the so-called ``WIMP miracle'': particles with weak-scale, {\it renormalizable} couplings to the SM and weak-scale masses can naturally account for the current relic density of DM.  If DM is a singlet under the SM gauge group, the only renormalizable coupling to SM particles is a quartic coupling of the SM Higgs boson and scalar DM \cite{Silveira:1985rk,McDonald:1993ex,Burgess:2000yq,Patt:2006fw,Barger:2007im}.  This model (which has been dubbed the ``Higgs portal'' model) has been well-studied in the literature.

In addition to the Higgs portal, it turns out that there are other renormalizable interactions between the SM and DM;  however, they require the introduction of additional fields besides the DM field.  The simplest of these is a cubic interaction of the form:
\bea
\Delta {\cal L} = g_{DM} \left( {\rm SM} \right) \left( \widetilde{{\rm SM}} \right) \left( {\rm DM} \right) \,,
\label{eq:EffDM1on1on1}
\eea
where $ \widetilde{{\rm SM}}$ is an additional field which, to preserve gauge invariance, must have the same quantum numbers as the SM field and is, thus, called a ``partner'' field.  As before, we assume that this interaction is invariant under a discrete symmetry which implies that both the DM field and the partner field are odd under this symmetry.  The DM particle is then completely stable as long as it is lighter than the partner particle (which we will assume in the following).

Recently, interactions of the form of Eq.~(\ref{eq:EffDM1on1on1}) where the SM field is a fermion have been the focus of several studies \cite{Chang:2013oia,Bai:2013iqa,DiFranzo:2013vra,Bai:2014osa,Chang:2014tea}.  These models (which  focused on interactions involving the first generation fermions) are interesting as they predict distinctive signals at both collider and direct detection experiments and, thus, are testable at current and near future experiments.

In this paper, we consider a variant of the ``quark portal'' model in which DM couples exclusively (or, at least, most strongly) to the top quark.  Similar models have recently been studied in Refs.~\cite{Kumar:2013hfa,Batell:2013zwa}\footnote{These types of models have also been studied in the context of neutrino mass generation \cite{Ng:2013xja,Ng:2014pqa}.}.  The motivation behind studying this scenario is another miracle of sorts which involves weak-scale couplings and masses:  the miracle of spontaneous breaking of the electroweak symmetry which has recently been confirmed with the discovery of a Higgs boson.  The apparent fact that both of these miracles involve weak-scale couplings and masses begs the question of whether or not electroweak symmetry breaking (EWSB) and the dynamics of DM are somehow related.  If they are, it might be reasonable to expect DM to have enhanced couplings to the most massive particles in the SM (e.g., the $W$ and $Z$ bosons, the Higgs boson and/or the top quark which is the case that we consider here).

In the following, we will consider a model where Dirac DM couples exclusively to the right-handed top quark via a Yukawa-like interaction involving a colored scalar.  The model is described in some detail in Section~\ref{sec:model}.   We then compute the relic density of DM (Section~\ref{sec:relic-density}) and map out the allowed parameter space by requiring our WIMP to account for the observed abundance.  Section~\ref{sec:direct-detect} contains an analysis of this type of dark matter at direct detection experiments.  In Section~\ref{sec:lhc}, we consider this model at the LHC where the dominant production channel will be $t\bar{t}$ plus missing energy.  We compare predictions for the signal event rates to those of the SM backgrounds and propose several kinematic variables which can discriminate between the signal and background.  Next, we turn our attention to investigating this model with indirect detection data, namely that from future gamma ray telescopes (Section~\ref{sec:indirect-det}).  We compute the continuum emission of gamma rays from WIMP annihilations as well as the loop-induced line emission and show that the resulting spectrum has very interesting and distinguishable characteristics.  Finally, we will conclude in Section~\ref{sec:conclusion}.

\section{The Top Portal Model}
\label{sec:model}

We consider a model where dark matter is made up entirely by a Dirac fermion ($\chi$) which couples exclusively to the right-handed top quark and is a singlet under the SM gauge group.  We call this type of dark matter ``Top Portal Dark Matter'' or TPDM.  In order to couple TPDM to the right-handed top quark in a renormalizable way, we also include a color-charged scalar ($\phi^a$) which, as required by gauge invariance, must have the same quantum numbers as the right-handed top quark.  The Lagrangian for the TPDM model is then given by:
\bea
{\cal L} = {\cal L}_{SM} + i \bar{\chi} \not{\partial} \chi - M_{\chi} \bar{\chi} \chi + \left( D_\mu \phi \right)^* \left( D^\mu \phi \right) - M_{\phi}^2 \phi^* \phi + \left( g_{DM} \phi^* \bar{\chi} t_R + h.c. \right) \,,
\eea
where the covariant derivative $D_\mu$ is given by:
\bea
D_\mu = \partial_\mu - i g_s G_\mu^a T^a - i Q_t e A_\mu + \frac{i e s_w}{c_w} Q_t Z_\mu \,.
\eea
The quantity $Q_t=+2/3$ is the charge of the top quark and $s_w (c_w)$ is the sine (cosine) of the weak-mixing angle $\theta_w$.  Here, we assume that $\phi$ does not obtain a mass via the SM Higgs mechanism for simplicity and, hence, there is no coupling between $\phi$ and the SM Higgs boson.

This model only has three free parameters: the coupling $g_{DM}$, the WIMP mass $M_\chi$ and the scalar partner mass $M_\phi$.  We require that the coupling remain perturbative ($g_{DM} < \sqrt{4 \pi}$) and, by construction, we assume that $M_\phi > M_\chi$ so that $\chi$ is completely stable against decay.  In the analysis that follows, we assume that the branching ratio of the decay $\phi \to \chi \bar{t}$ is 100\% and the width of the $\phi$ particle is:
\bea
\Gamma(\phi \to \chi + \bar{t}) = \frac{g_{DM}^2}{16\pi} \frac{ \left( M_\phi^2 - M_\chi^2 - m_t^2 \right)}{M_\phi^3} \sqrt{M_\phi^4 - 2 M_\phi^2 \left(M_\chi^2 + m_t^2 \right) + \left(M_\chi^2 - m_t^2 \right)^2} \,.
\label{eq:phi-width}
\eea
%

\section{The Dark Matter Relic Abundance}
\label{sec:relic-density}

\begin{figure}[t]
\begin{center}
\includegraphics[scale=0.26]{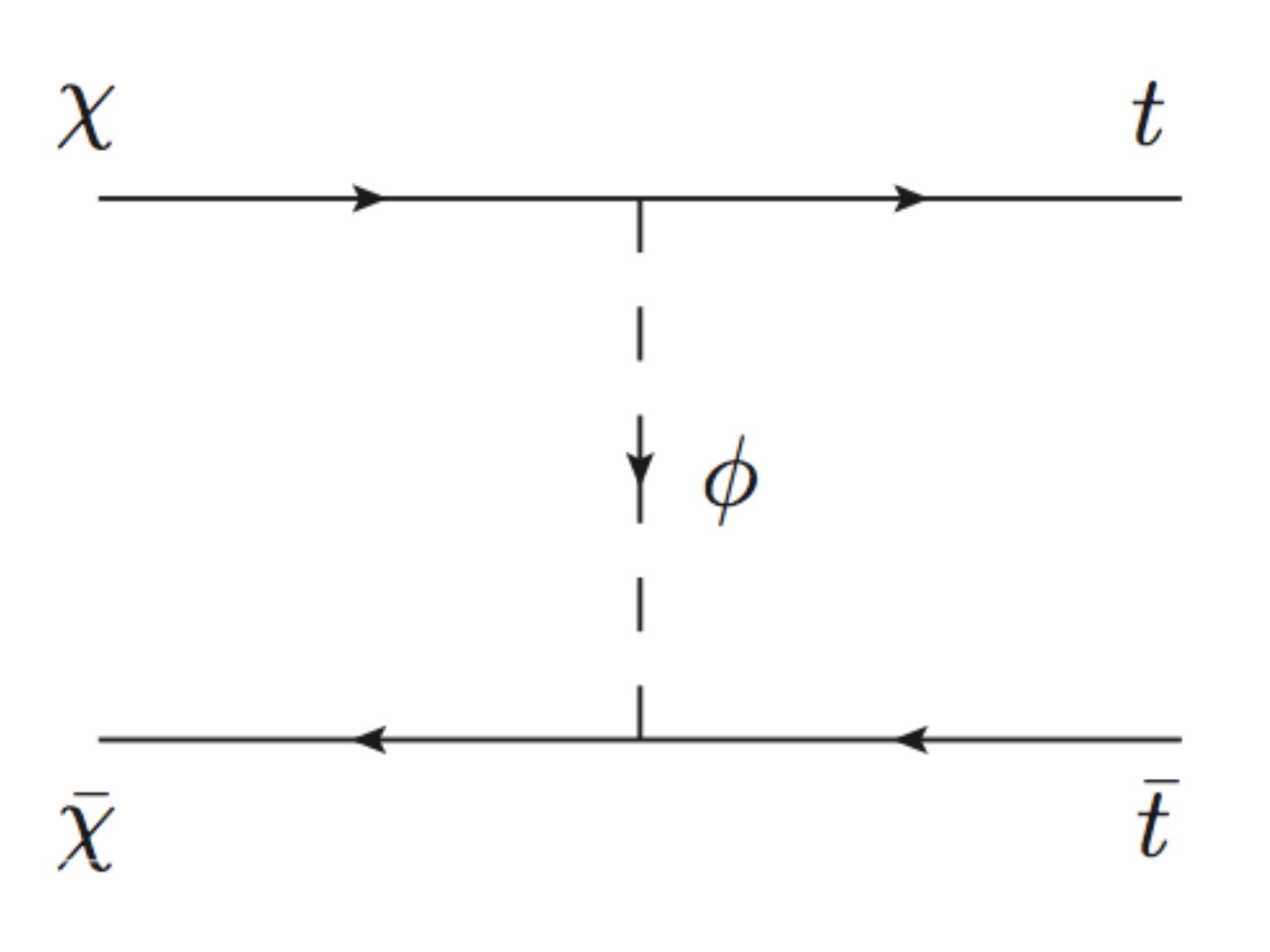}
\hspace{0.50cm}
\includegraphics[scale=0.26]{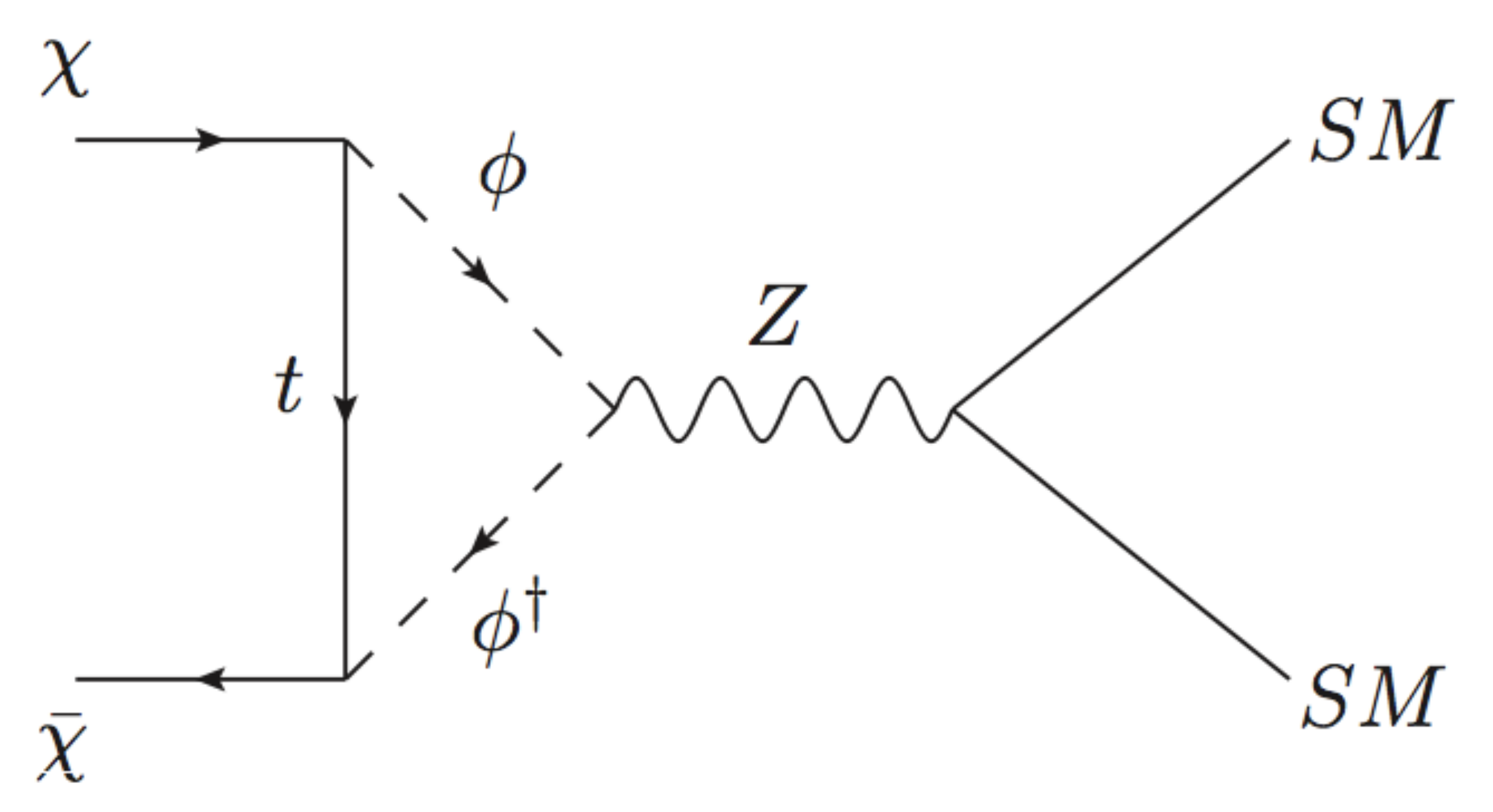} 
\caption{Feynman diagrams contributing to dark matter freeze-out.  The first diagram represents the leading annihilation mode, while the second diagram is an example of a loop-level process which allows the WIMPs to annihilate into light SM states.}
\label{fig:annihilation}
\end{center}
\end{figure}

In this section, we consider the constraints on TPDM from relic density measurements.  To begin, we compute the annihilation cross section analytically and derive constraints on TPDM closely following the approach in Ref.~\cite{Bai:2013iqa}.  Annihilation of dark matter in the early universe in the TPDM model occurs via $t$-channel $\phi$ exchange and results in a final state of $t\bar{t}$ as depicted in the first Feynman diagram of Fig.~\ref{fig:annihilation}.  In addition, since $\phi$ carries the SM quantum numbers of the right-handed top quark, loop-level annihilations into lighter SM states are also possible (but, typically suppressed compared to the tree-level process).  Because the WIMPs are assumed to be non-relativistic, the annihilation cross section can be computed as an expansion in the WIMP velocity $\beta$, where $\beta \simeq 0.3$ near freeze-out.  Keeping terms up to ${\cal O}(\beta^2)$ we find:
\bea
 \left( \sigma v \right)_{t\bar{t}} &=& \frac{3 g_{DM}^4 \left(2 M_\chi^2 + m_t^2 \right)^2}{256 \pi M_\chi^2 \left(M_\chi^2 + M_\phi^2 \right)^2} - \frac{g_{DM}^4 }{256 \pi M_\chi^2 \left(M_\chi^2 + M_\phi^2 \right)^4} \biggl[ 16 M_\chi^4 \left(M_\chi^4 + 3 M_\chi^2 M_\phi^2 - M_\phi^4 \right) \nonumber\\
&&\,\,\,\,\,\,\,\, + 16 m_t^2 M_\chi^4 \left( M_\chi^2 + 4 M_\phi^2 \right) - 3 m_t^4 \left( M_\chi^4 + 6 M_\chi^2 M_\phi^2 + M_\phi^4 \right) \biggr]  \, \beta^2  \nonumber\\
&\equiv& s + p \beta^2  \,.
\eea

\begin{figure}[t]
\begin{center}
\includegraphics[scale=0.26]{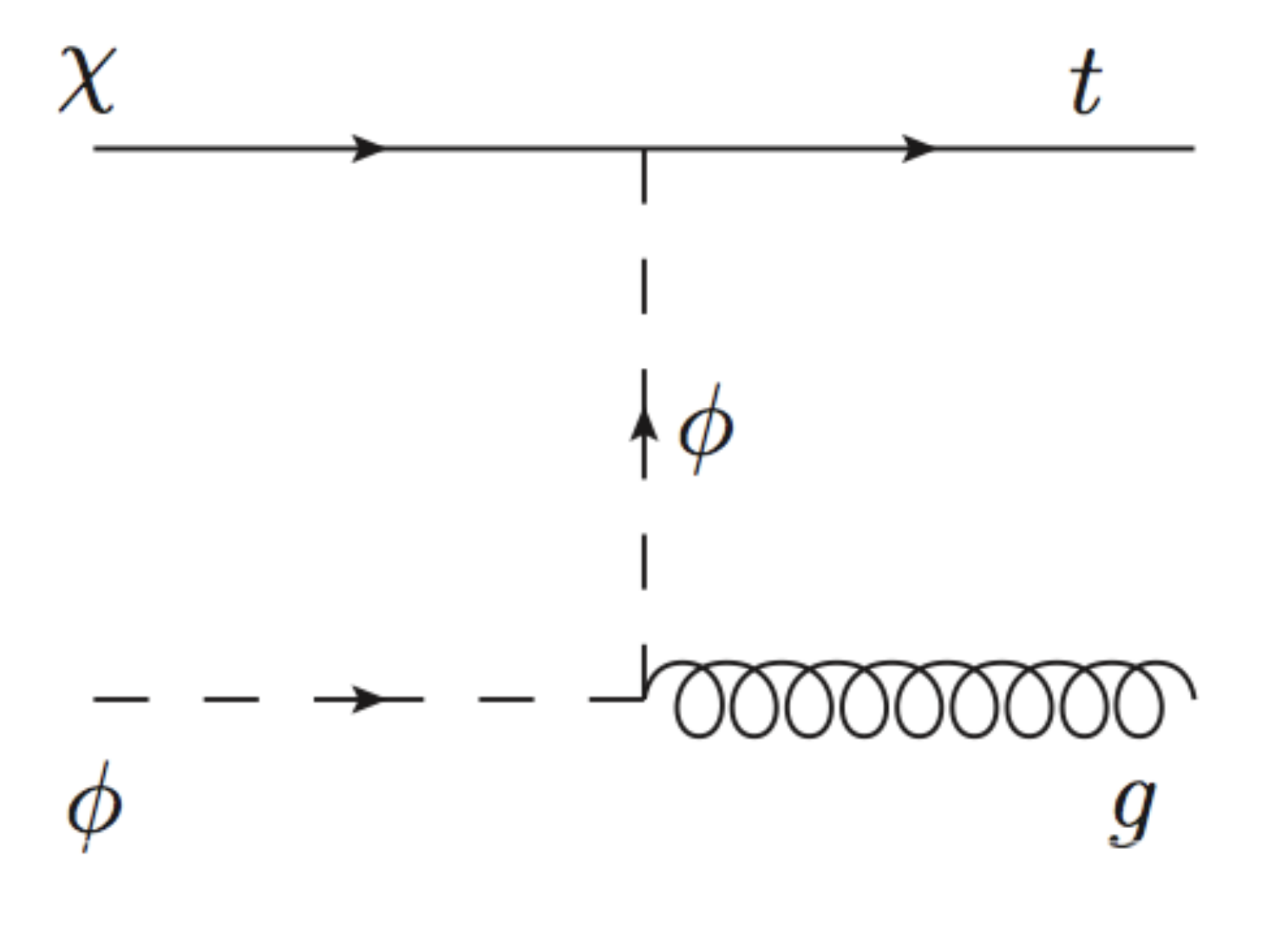}
\hspace{1cm}
\includegraphics[scale=0.2]{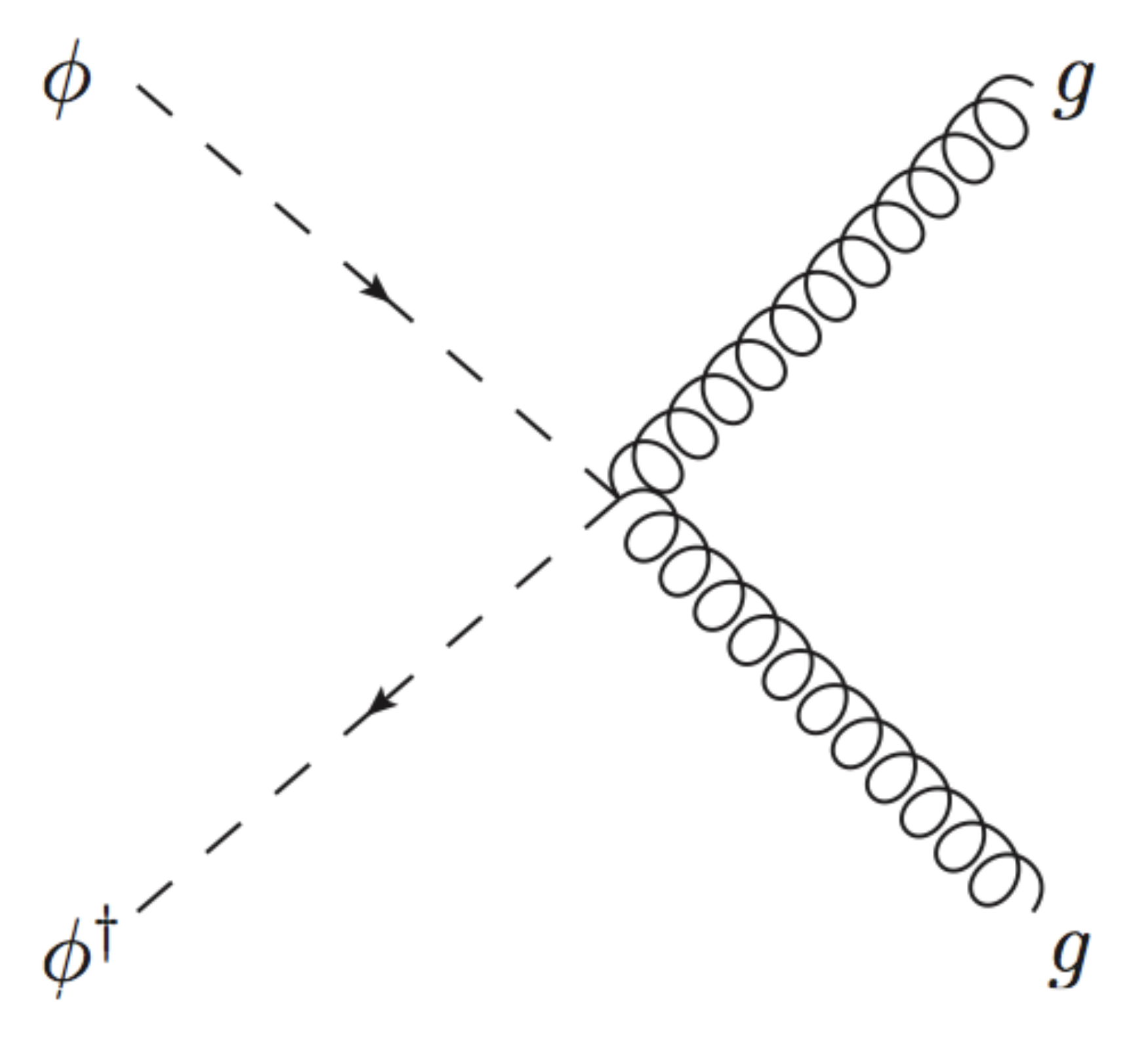}
\caption{Example Feynman diagrams of co-annihilation modes which become important for the relic density when $M_\phi \simeq M_\chi$.}
\label{fig:coannihilation}
\end{center}
\end{figure}
The $s$ and $p$ variables determine the relic density via the approximation that is valid away from threshold:
\bea
\Omega_\chi h^2 \approx \frac{1.07 \times 10^9}{{\rm GeV} \, M_{Pl} \sqrt{g^*}} \frac{x_F}{s + 3p/x_F},
\label{eq:Omega}
\eea
where $M_{Pl}=1.2\times 10^{19}$ GeV is the Planck scale and $g^*$ is the relativistic degrees of freedom present at the time of freeze-out ($g^* = 86.25$).  The freeze-out temperature ($x_F$) is given by the implicit transcendental equation:
\bea
x_F = \ln \left[ \frac{5}{4} \sqrt{\frac{45}{8}} \frac{g}{2\pi^3} \frac{M_{Pl} M_\chi (s + 6 p/x_F)}{\sqrt{g^*} \sqrt{x_F}} \right] \,,
\label{eq:xF}
\eea
where $g$ represents the number of degrees of freedom for the WIMP ($g = 4$ in our case).  By solving the coupled equations (Eq.~(\ref{eq:Omega}) and (\ref{eq:xF})), we can eliminate one of the model's free parameters by requiring the relic density match the measured value.  For example, we can determine the coupling needed for the set of $M_\chi$ and $M_\phi$ values.

In regions of parameter space where the $\phi$ and $\chi$ masses are non-degenerate, the above analysis is sufficient.  However, in regions where the mass splitting is small, co-annihilation effects are important.  In these cases, the relic density is determined from a number of processes whose relative contributions depend principally on the $M_\phi-M_\chi$ splitting and other parameters of the model.  Some examples of co-annihilation processes include $\phi  \phi^\dagger \to g g$, $\chi \phi^\dagger \to t g$, $\phi  \phi^\dagger \to \gamma g$ and $\chi \phi^\dagger \to b W^+$, a couple of which are depicted in Fig.~\ref{fig:coannihilation}.

\begin{figure}[t]
\begin{center}
\includegraphics[scale=0.60]{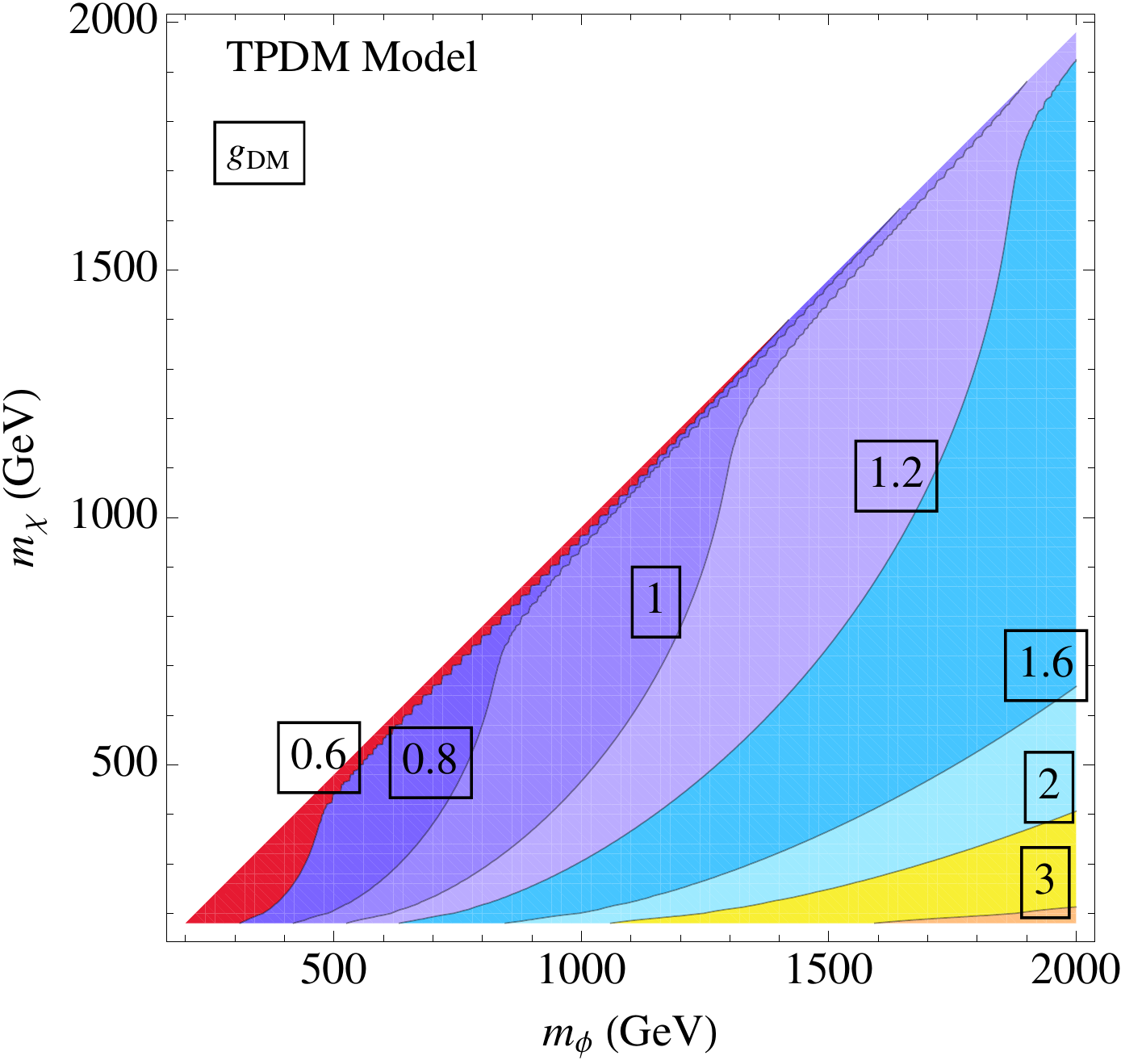} 
\caption{The WIMP and scalar partner masses (for various couplings) which result in the correct relic density of dark matteras measured by WMAP \cite{Bennett:2012zja} and Planck \cite{Ade:2013zuv}. }
\label{fig:relic-density-tR}
\end{center}
\end{figure}

In regions of co-annihilation, the analytic approach described is insufficient.  We have therefore implemented the model in micrOMEGAs \cite{Belanger:2013oya} which includes all co-annihilation processes.  We have checked that, in regions neglecting co-annihilation, the two approaches agree for the prediction of the relic density for various values of the model parameters.  The results for the fit to the measured value of the relic density from WMAP \cite{Bennett:2012zja} and Planck \cite{Ade:2013zuv} with
\be
\Omega h^2 = 0.1199 \pm 0.0027
\ee
are shown in the left panel of Fig.~\ref{fig:relic-density-tR}.  In the plot, we show contours in the $M_\phi - M_\chi$ plane for various couplings $g_{DM}$ which satisfy the relic density constraint.  The importance of co-annihilation effects is evident in the long tails of the contours in the $M_\chi \to M_\phi$ limit.

In the following analyses of TPDM at the LHC and indirect detection of TPDM via gamma rays, we will only consider values of the model parameters ($g_{DM}, M_\chi$ and $M_\phi$) which give the correct relic abundance.

\section{TPDM at Direct Detection Experiments}
\label{sec:direct-detect}

\begin{figure}[t]
\begin{center}
\includegraphics[scale=0.26]{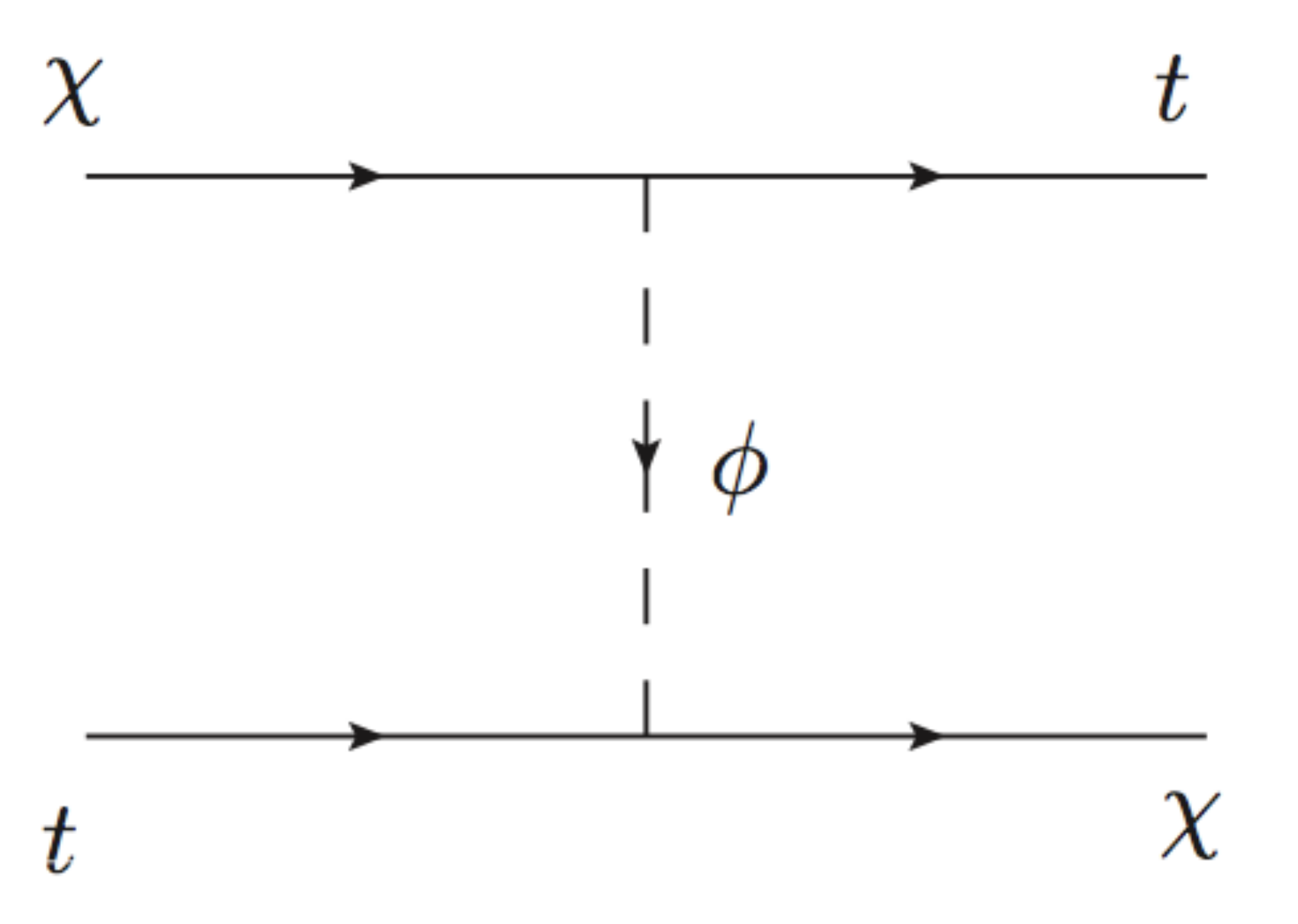}
\hspace{0.5cm}
\includegraphics[scale=0.23]{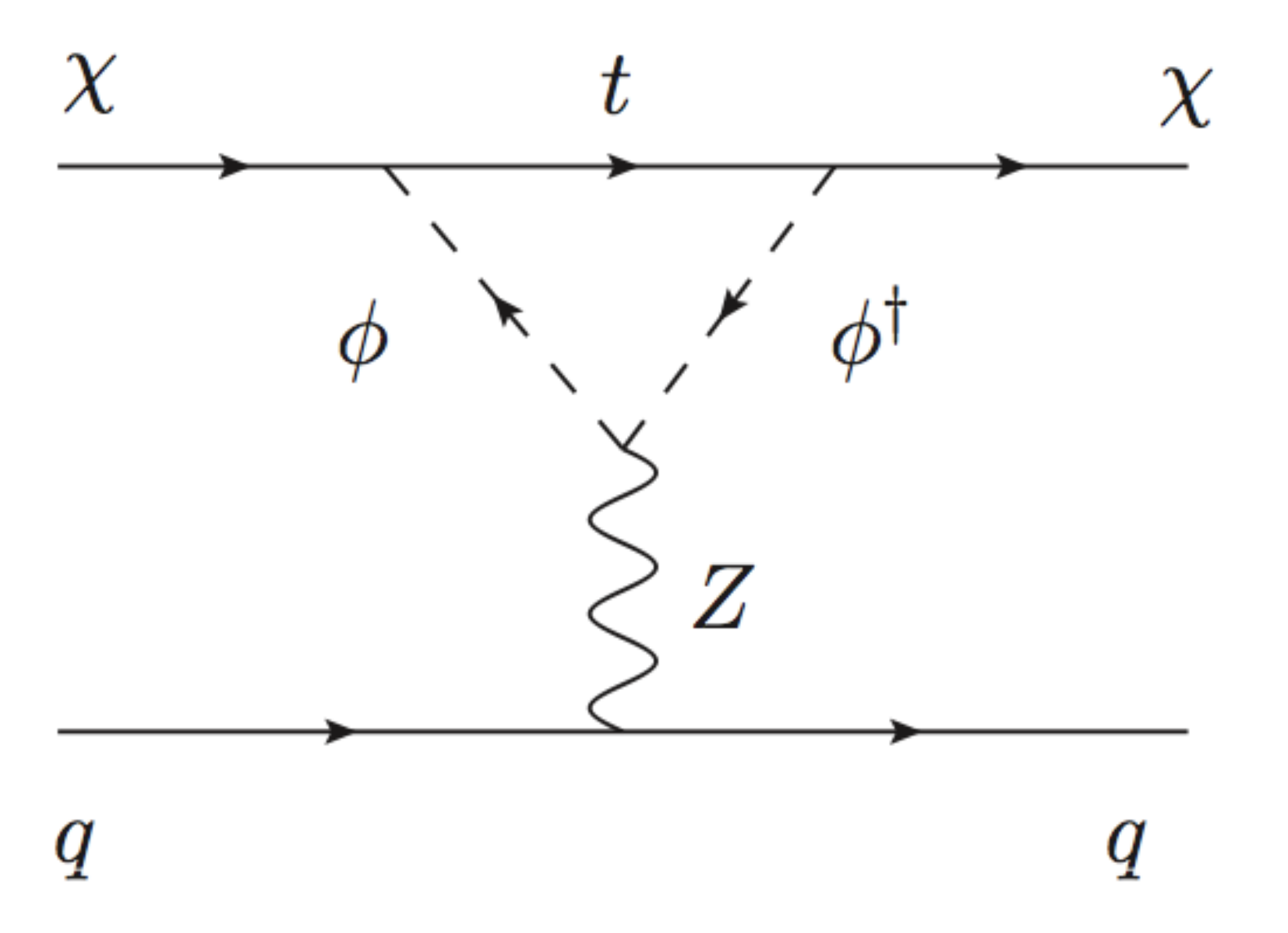}
\hspace{0.5cm}
\includegraphics[scale=0.23]{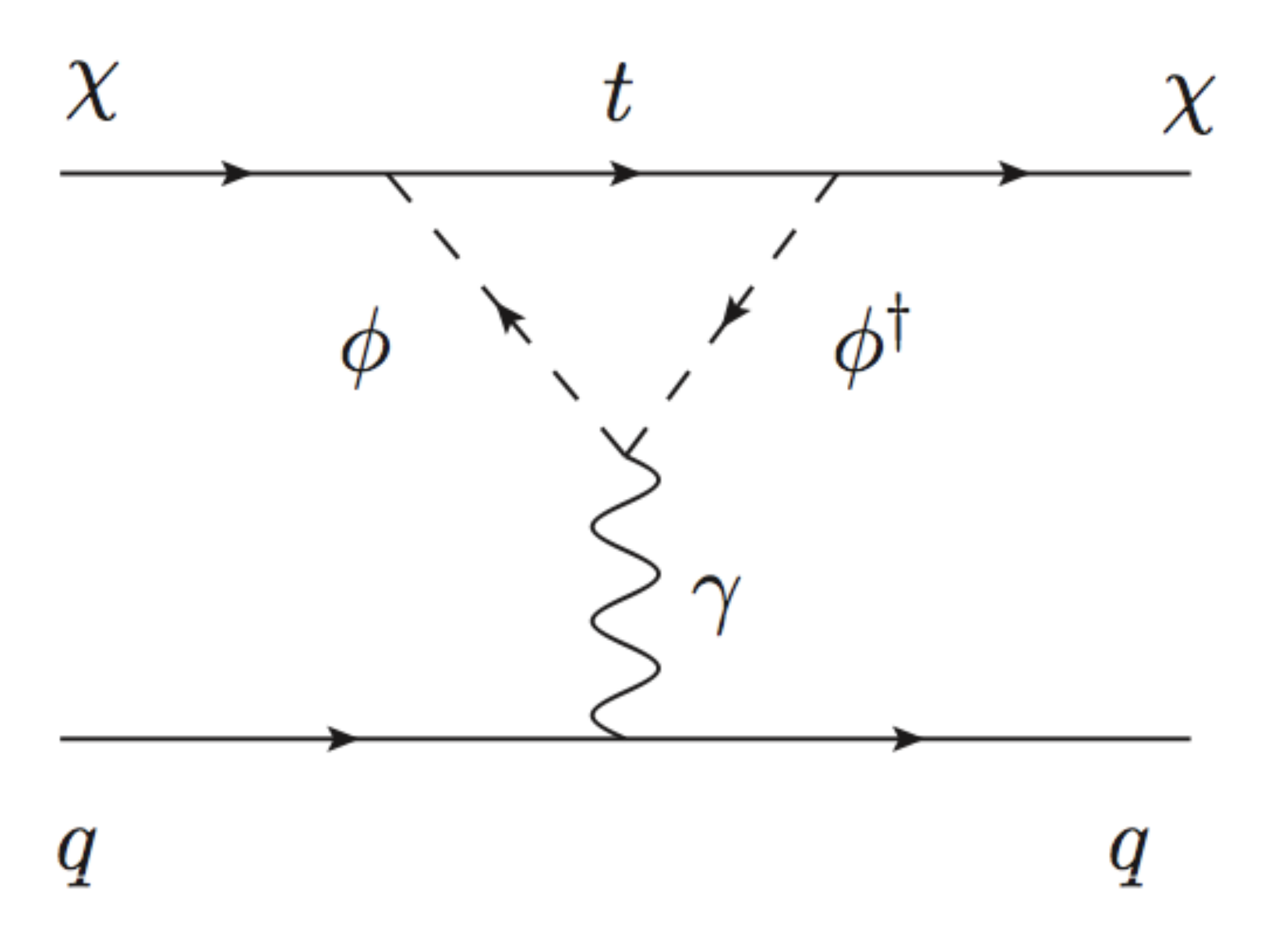}
\caption{Feynman diagrams contributing to direct detection of TPDM.  The leading-order diagram (left) is highly suppressed due to the smallness of the top quark PDF.  The first non-vanishing contribution would come from loop-level diagrams, an example of which is presented in the right two panels.  The $\gamma$ exchange contributes to the magnetic dipole and charge-dipole interactions which exhibit a nontrivial $E_r$ dependence.}
\label{fig:dd}
\end{center}
\end{figure}

Constraints from direct detection experiments are nonexistent at the tree-level due to the top quark parton distribution function in the proton, (c.f. Fig.~\ref{fig:dd}a).  However, at one loop, additional diagrams yield a sizable contribution to the scattering rate.  These include the $t$-channel $Z$ exchange, whose differential recoil energy, $E_r$, distribution takes the form~\cite{Batell:2013zwa}
\be
{d \sigma_Z\over d E_r} = {m_T\over 2 \pi v^2} (f_p Z + (A-Z) f_n)^2 F(E_r)^2,
\ee
where $m_T$ is the mass of the target, which in the cases of LUX and Xenon100 is Xe.  The nuclear form factor, denoted $F(E_r)$, that we adopt is the Woods-Saxon Form Factor~\cite{Engel:1991wq}.  The effective neutron and proton couplings are given by
\be
f_n = {g_Z G_F c_w \over \sqrt 2 g}, \quad \quad f_p = {(4 s_w^2-1) g_Z G_F c_w\over \sqrt 2 g},
\ee
The $\bar \chi \chi Z$ coupling, $g_Z$, is to leading order in the large $m_\phi$ limit given by
\be
g_Z \approx {N_c g g_{DM}^2 m_t^2 \over 16 \pi^2 c_w m_\phi^2} \left( 1+ \log\left({m_t^2\over m_\phi^2}\right)\right).
\ee

Additional contributions include the magnetic dipole and the charge-charge contribution.  The magnetic dipole contribution is given by
\be
{d \sigma_{DZ}\over d E_r} = {e^2 Z^2 \mu_\chi^2 \over 4 \pi v^2 m_T} \left( {m_T v^2\over 2 E_r} - {M_\chi+2 m_T\over 2 M_\chi}\right) F(E_r)^2,
\ee
where $\mu_\chi\approx {e g_{DM}^2 M_\chi\over 32 \pi^2 M_\phi^2}$ is the magnetic dipole in the large $m_\phi$ limit.  We use the full one-loop result and additional charge-charge interaction found in Ref.~\cite{Batell:2013zwa}.  

To apply these interactions to the presently known scattering cross section limits, we follow the method outlined in Ref.~\cite{Batell:2013zwa}.  Namely, due to the nontrivial dependence on recoil energy in the magnetic dipole interaction, we integrate the recoil energies over the range $5-25~ {\rm keV}$ range and require two events to arrive at the limit given by LUX with a $10^4$ kg d exposure.  We have independently verified that this procedure yields a limit comparable with the LUX collaboration.

\begin{figure}[t]
\begin{center}
\includegraphics[width=0.49\textwidth]{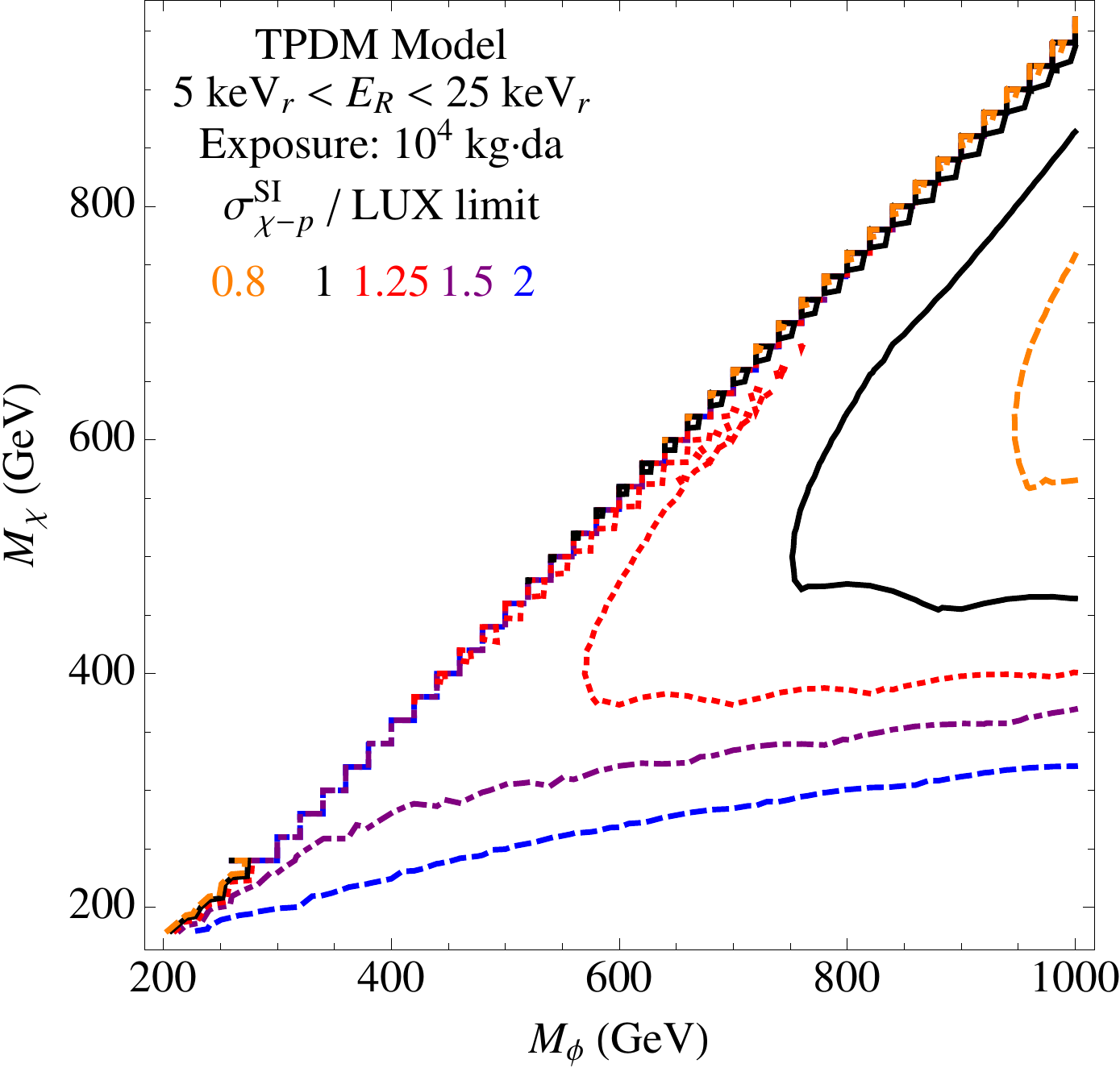}
\caption{Limits from the LUX 85da measurement.  Contours are shown in selected ratios of the predicted scattering cross section to the published LUX result, with the solid curve indicating the LUX limit.}
\label{fig:ddresult}
\end{center}
\end{figure}

We find the $Z$ exchange contribution is typically dominant over the parameter space we consider.   We show in Fig.~\ref{fig:ddresult}, the scattering constraints from LUX.  These limits indicate a region of $M_\chi > 450$ GeV and $M_\phi > 750$ GeV is required to be in agreement with LUX.   However, the uncertainties in detection efficiency and the position of the nuclear recoil band yield up to a 25\% $1\sigma$ uncertainty in the location of the limit.   This is indicated in Fig.~\ref{fig:ddresult} by the red dotted line, which permits a region of roughly $M_\chi > 400$ GeV and $M_\phi > 600$ GeV.  Furthermore, we remark that uncertainties in the local dark matter density and velocity can further loosen this constraint.  Additionally, WIMP dark matter may have multiple components, yielding a lower local density than expected.  Taken together, these limits should serve as a rough guide, rather than a hard rule.

\section{LHC Prospects}
\label{sec:lhc}

\begin{figure}[htbp]
\begin{center}
\includegraphics[scale=0.2]{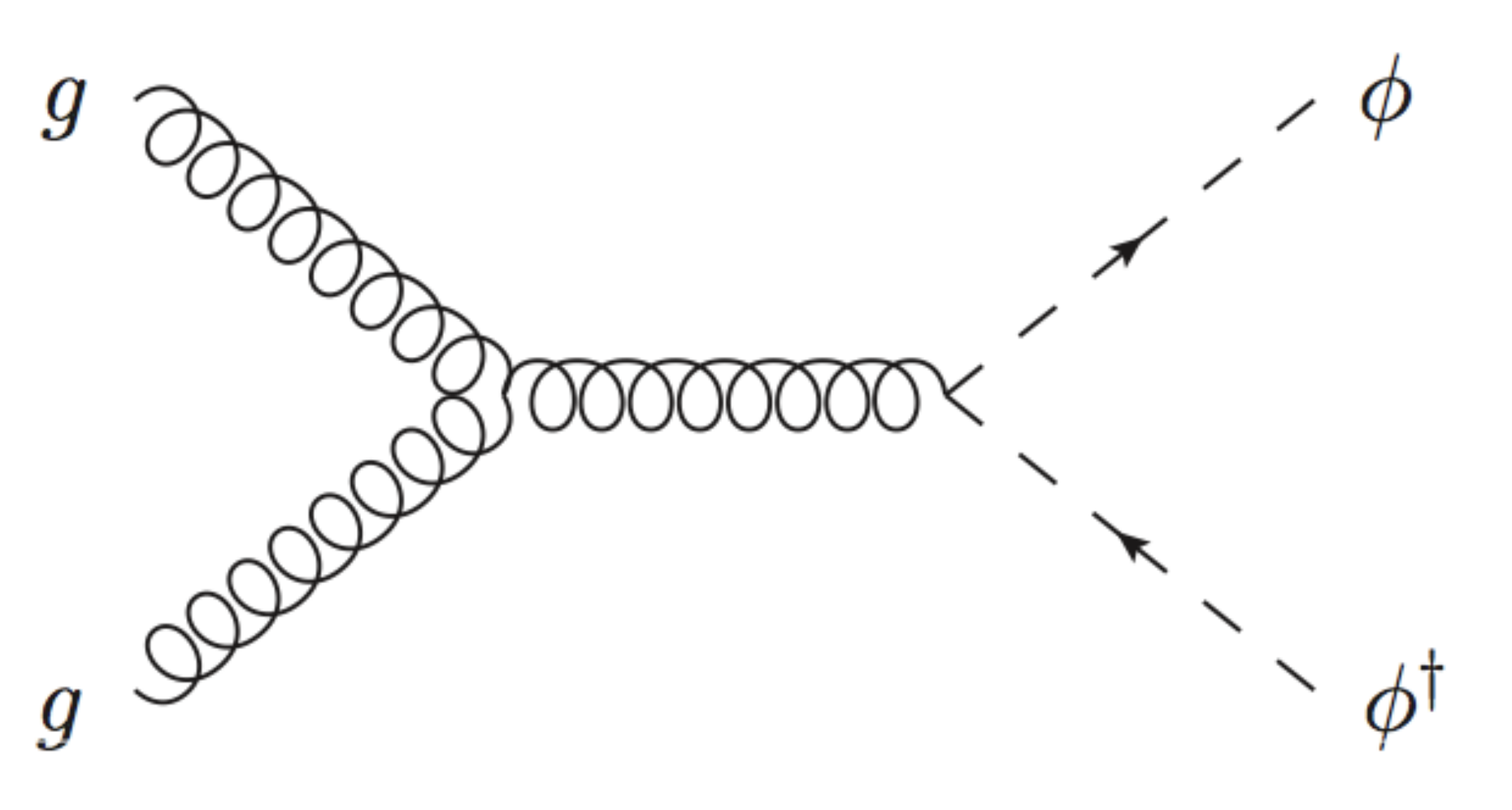}
\hspace{0.50cm}
\includegraphics[scale=0.2]{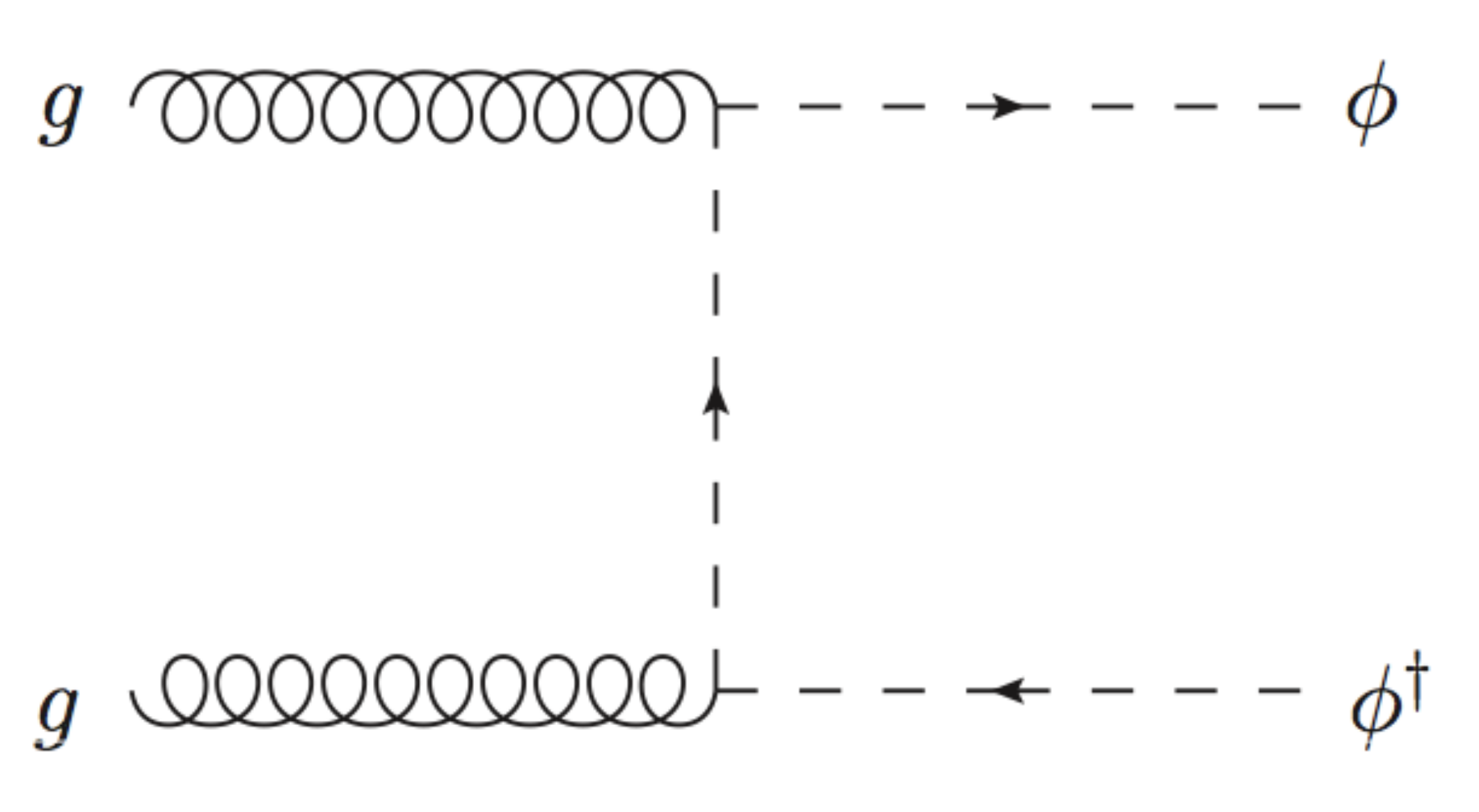}
\hspace{0.50cm}
\includegraphics[scale=0.2]{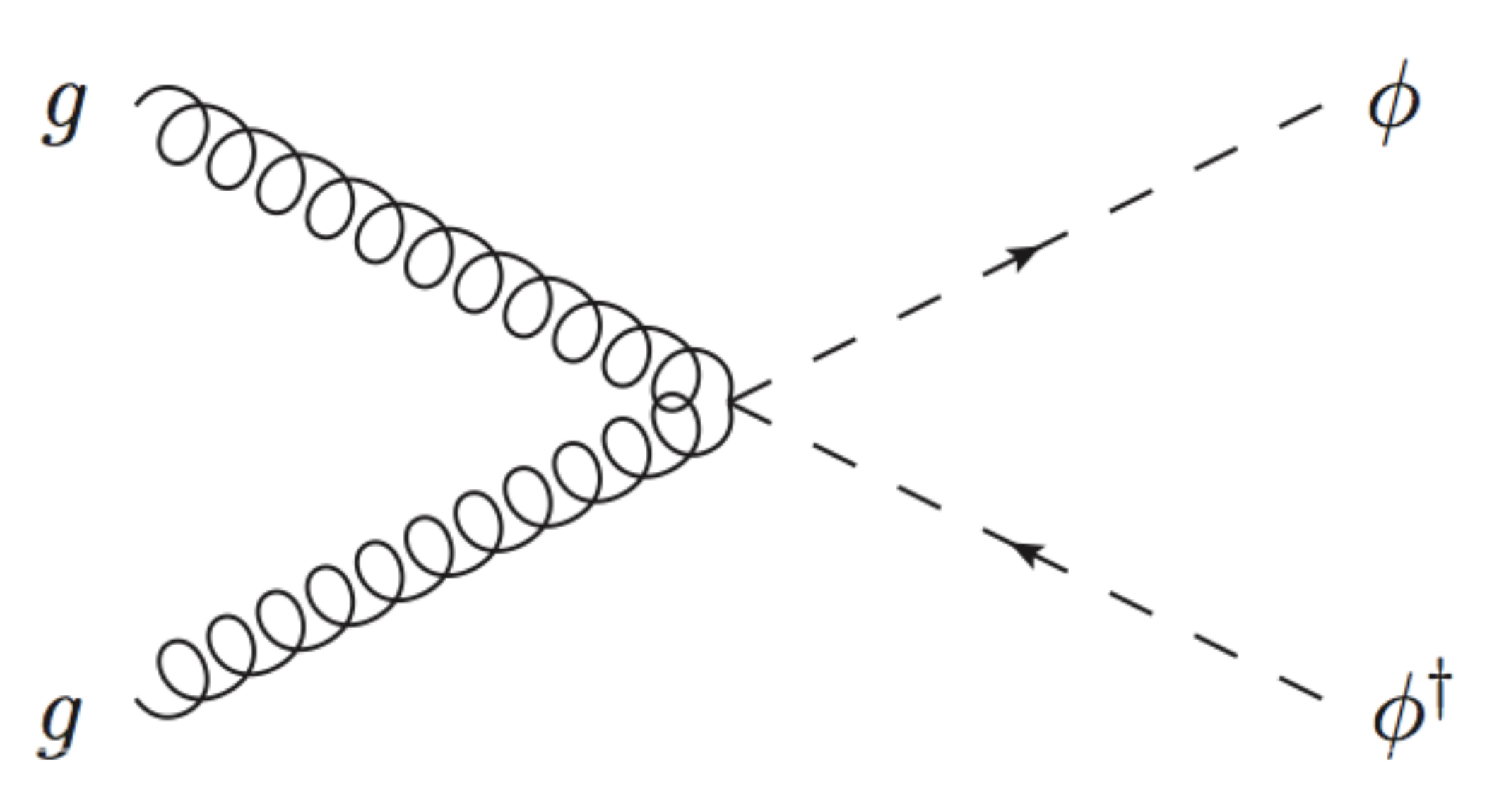}
\caption{Feynman diagrams which contribute to the production of TPDM at the LHC. }
\label{fig:LHCprod}
\end{center}
\end{figure}

The production of TPDM at the LHC occurs through the pair production of the colored scalars as depicted in Fig.~\ref{fig:LHCprod}.  Since the scalars are $SU(3)_c$ triplets, they are produced in a similar manneer as to the scalar quarks found in SUSY models.  However, unlike SUSY models, our scenario does not include a chargino component.  Therefore, most SUSY search analyses for $\tilde t \to t \chi^0_1$ cannot be mapped to our scenario as they include light charginos.  A model independent analysis is available, but with restricted $M_\chi$ and limited luminosity~\cite{Aad:2011wc}.

To gauge the prospects for discovering this scenario with future LHC Run-II data, we perform collider simulations of the signal
\be
pp \to \phi \phi^* \to t \bar \chi + \bar t \chi \to b\bar b \ell^\pm j j + \met,
\ee
using Madgraph-5~\cite{Alwall:2011uj}.  We note that since the dominant decay mode of the $\phi$ is to $t \bar \chi$, in the narrow width approximation, the production cross section of the $t\bar t \chi \bar \chi$ final state is dependent only on the production of $\phi$.  Hence, the value of  $M_\phi$ (and $g_s$, the strong coupling) completely determine the production cross section. 

Similarly, we simulate the pertinent backgrounds and find that the SM 
\be
pp\to t\bar t \to b\bar bj j \ell \nu  
\ee
background is dominant.  

To account for $b$-jet tagging efficiencies, we assume a $b$-tagging rate of 70\% for $b$-quarks with $p_T > 30~{\rm GeV}$ and $|\eta_{b}| < 2.4$ consistent with multivariate tagging suggested for the LHC luminosity upgrade~\cite{atlphyspub2013004}.  We also apply a mis-tagging rate for charm-quarks as:
\begin{equation}
\epsilon_{c\to b} =10\% \quad\quad {\rm for } \quad p_T(c) > 50 {\rm GeV},
\end{equation}
while the mis-tagging rate for a light quark or gluon is: 
\begin{eqnarray}
\epsilon_{u,d,s,g\to b} &= 2\% \quad\quad \quad {\rm for }& p_T(j)> 250 {\rm GeV},\\
\epsilon_{u,d,s,g\to b} &= 0.67\% \quad\quad {\rm for }& p_T(j) < 100 {\rm GeV}.
\end{eqnarray}
Over the range $100{\rm GeV}<p_T(j)<250{\rm GeV}$, we linearly interpolate the fake rates given above~\cite{Baer:2007ya}.  With pile-up the rejection rate is expected to worsen by up to 20\%~\cite{atlphyspub2013004}. Finally, we model detector resolution effects by smearing the final state energy according to:
\begin{equation}
{\delta E \over E} = {a \over \sqrt{E}} \oplus b,
\end{equation}
where we take $a=50\%$ and $b=3\%$ for jets and $a=10\%$ and $b=0.7\%$ for photons.

We apply both a cut-based analysis and a multi-variate analysis (MVA) which relies on relevant kinematic variables.  For either case, we require the following tags:
\be
n_b^{\rm tag} = 2, \quad \quad n_j^{\rm tag} = 2, \quad \quad n_\ell^{\rm tag} = 1.
\label{eqn:tag}
\ee
For the cut-based analysis, we apply cuts on  $\Delta R_{ab}=\sqrt{(\phi_a-\phi_b)^2+(\eta_a-\eta_b)^2}$, the separation of two objects in the $\eta-\phi$ plane. The cuts applied are 
\bea
\Delta R_{jj, b\bar b, b j} &>& 0.4,\quad \quad\qquad \Delta R_{j\ell, b\ell} > 0.2,\\
p_T(j)&>& 25~{\rm GeV},\quad \quad\quad |\eta_{j}| < 2.4 ,\\
p_T(e,\mu)&>& 25, 20~{\rm GeV},\quad |\eta_{e,\mu}| < 2.5 .
\eea

The main distinguishing aspect of the $\phi \phi$ signal is the strong missing energy signature.  We find that a beginning cut of 
\be
\met > 100~{\rm GeV}
\ee
retains a majority of the signal while removing a bulk of the background.  

\begin{figure}[htbp]
\begin{center}
\includegraphics[width=0.3\textwidth]{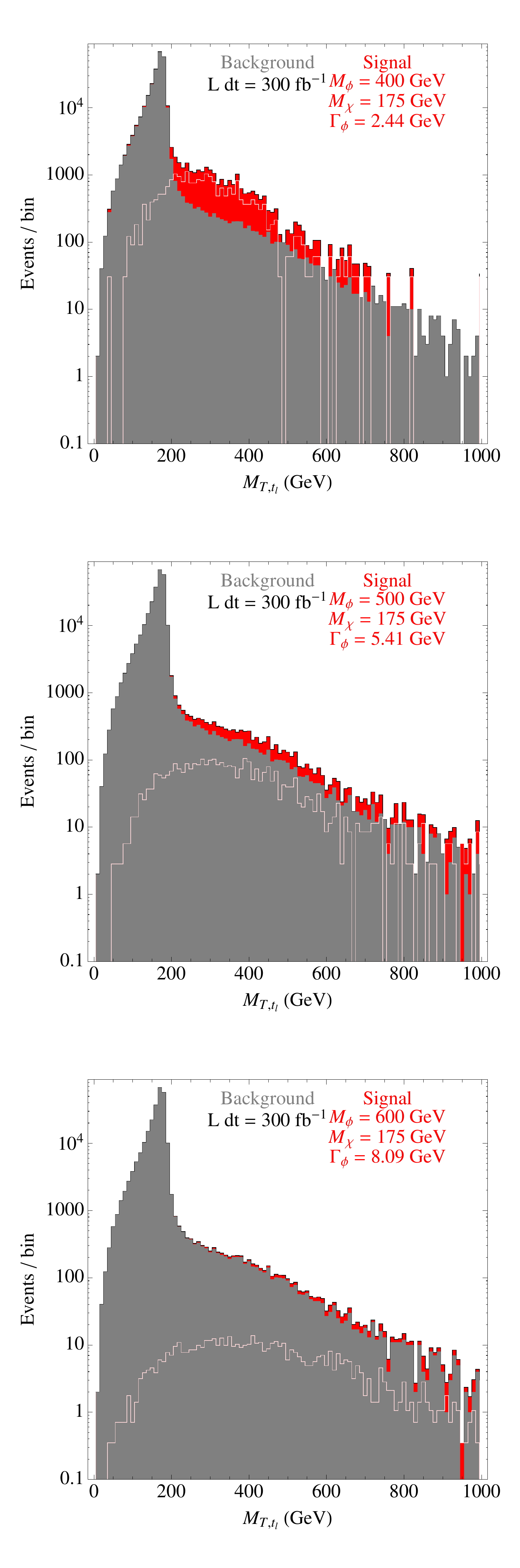} 
\includegraphics[width=0.3\textwidth]{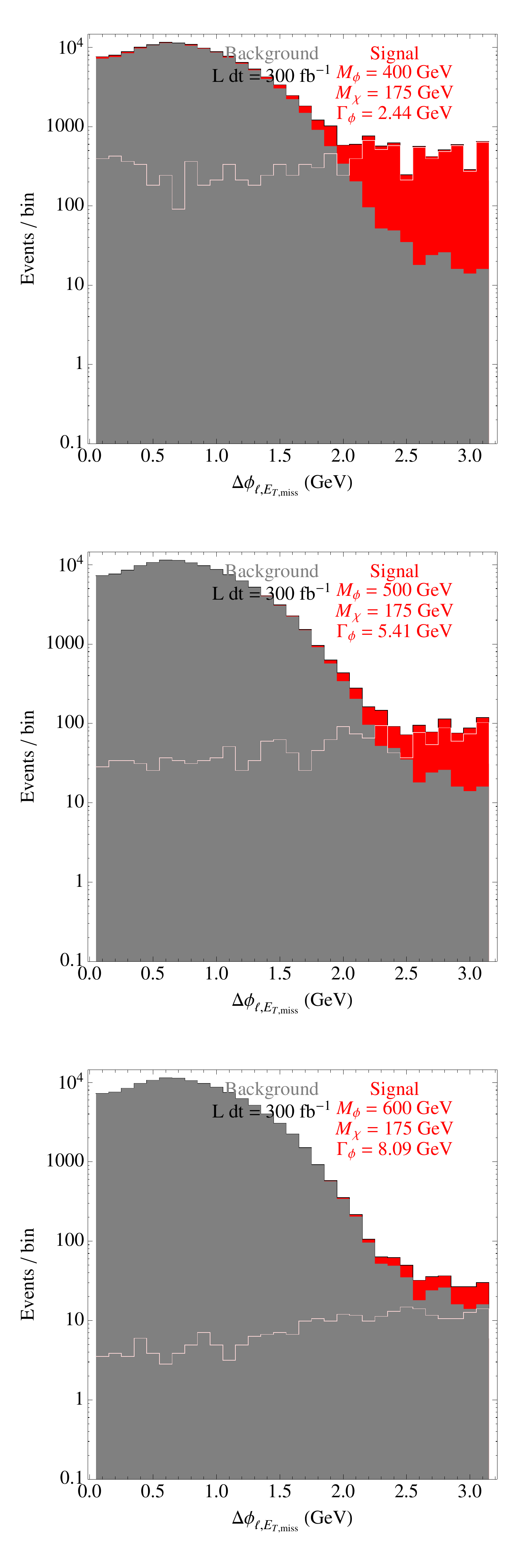} 
\includegraphics[width=0.3\textwidth]{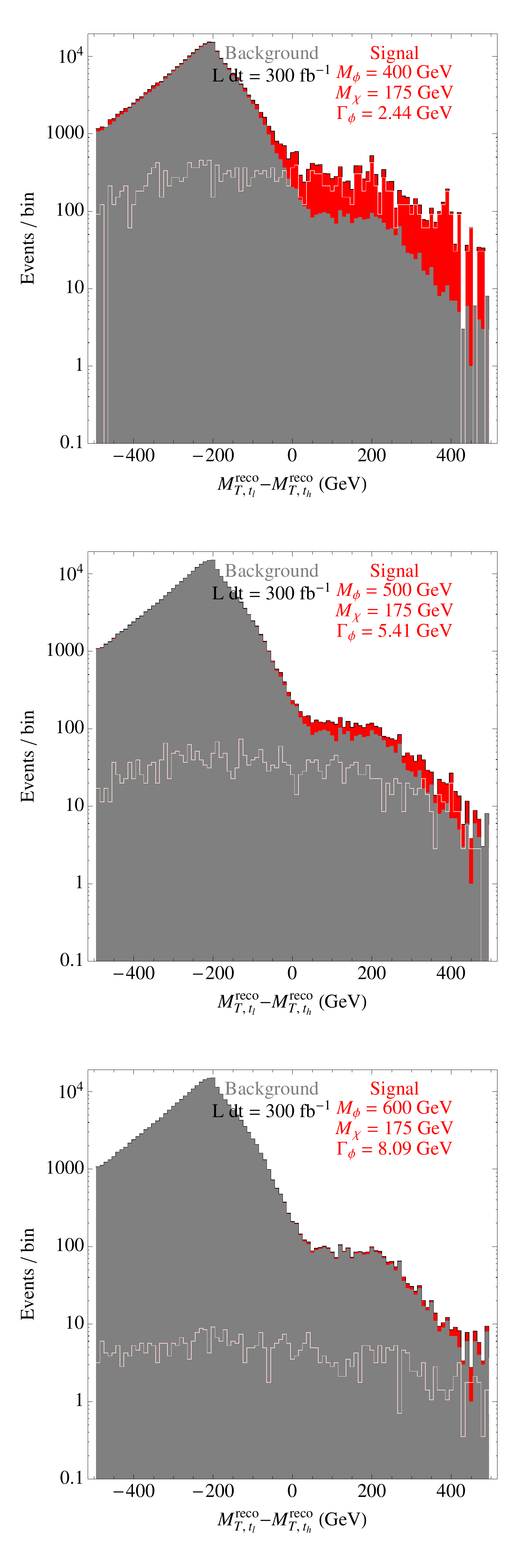} 
\caption{Signal and background distributions after the initial cuts defined in Eqs.~\ref{eqn:tag}-\ref{eqn:mtreco} of the (left panels) $M_T(b\ell,\met)$, (middle panels) $\Delta \phi_{\ell,~\met}$ and (right panels) $M_T(\phi_h) - M_T(\phi_\ell) $ for $M_\phi = 400, 500 $ and 600 GeV.  }
\label{fig:dists}
\end{center}
\end{figure}

We then require that one tagged $b$-jet and two additional jets reconstruct the hadronically decaying top 
\be
|M_{b jj} - m_t| < 20~{\rm GeV},
\label{eqn:mtreco}
\ee
and note the other $b$-tag and charged lepton originate from the other, leptonically decaying top quark.  The transverse cluster mass $M_T(b\ell,\met)$ is a suitable variable for this side of the decay~\cite{Barger:1987re}:
\be
M_T(a,b) = (|p_T(a)| + |p_T(b)|)^2 - (p_T(a) + p_T(b))^2,
\ee
where the observable cluster is $a=b\ell$.  This variable generically has an upper bound which is related to the mass of the parent particle, $M_T(a,b) \le M_{ab}$.  Since the typical $\met$ distribution from a SM $t$-quark decay is bounded by $m_t$, we find the signal can be isolated if it has an appreciable cross section above $M_T(b\ell,\met) > 200$ GeV.  This is illustrated in the first column of Fig.~\ref{fig:dists} for selected points in parameter space.

Another strong discriminator is the azimuthal angle between $\ell$ and $\met$.  In the $t\bar t$ SM background, the leptonically decaying $W$-boson does not often decay near rest as it is boosted from the top quark decay.  Therefore, the $\ell$ and $\nu$ directions are correlated and close together.  This is contrasted with the signal topology in which the $\ell$ is paired with the $\nu$ from the $W$-decay as in the $t\bar t$ background, but the additional $\chi \bar \chi$ system disrupts this correlation.  Therefore, we expect to see a separation among the signal and background in this $\Delta \phi_{\ell,~\met}$ observable.  Indeed, this is the case and can be seen in the middle column of Fig.~\ref{fig:dists}.

Furthermore, in addition to the~~$\met$, $M_T$ and $\Delta \phi_{\ell,~\met}$  cuts, we observe that the $\phi$ pair is produced back-to-back.  Therefore, the total transverse momentum carried off in either side of the $\phi$ decay should be balanced~\footnote{Assuming the transverse CM frame is in the lab frame.  A boost to the CM frame can be made if ISR kicks the CM frame in the transverse direction.}
\be
p_T(\phi_h) = - p_T(\phi_\ell),
\ee
where we denote $\phi_{h}$ and $\phi_{\ell}$ as the $\phi$ which decays through a hadronically and leptonically decaying top quark, respectively.  Therefore, one can relate the $\met$ from both sides of the decay 
\bea
\met^{\ell} &=& {1\over 2} (\met^{\rm obs} + p_T(t_h) - p_T(b \ell)),\\
\met^{h} &=& {1\over 2} (\met^{\rm obs} - p_T(t_h) + p_T(b \ell)),
\eea
where $\met^{\ell}$ and $\met^h$ denote the missing energy from the leptonic and hadronic sides of the $\phi$ decays, and $\met^{\rm obs}$ is the observed missing energy.  The $M_T$ variable applied to  both sides of the event offers additional discrimination power.  Specifically, the difference between the leptonic and hadronically decaying side, $M_T(\phi_h) - M_T(\phi_\ell) $, shows a modest separation in the signal and background, enough to provide an additional check of the signal.  

\begin{figure}[htbp]
\begin{center}
\includegraphics[width=0.49\textwidth]{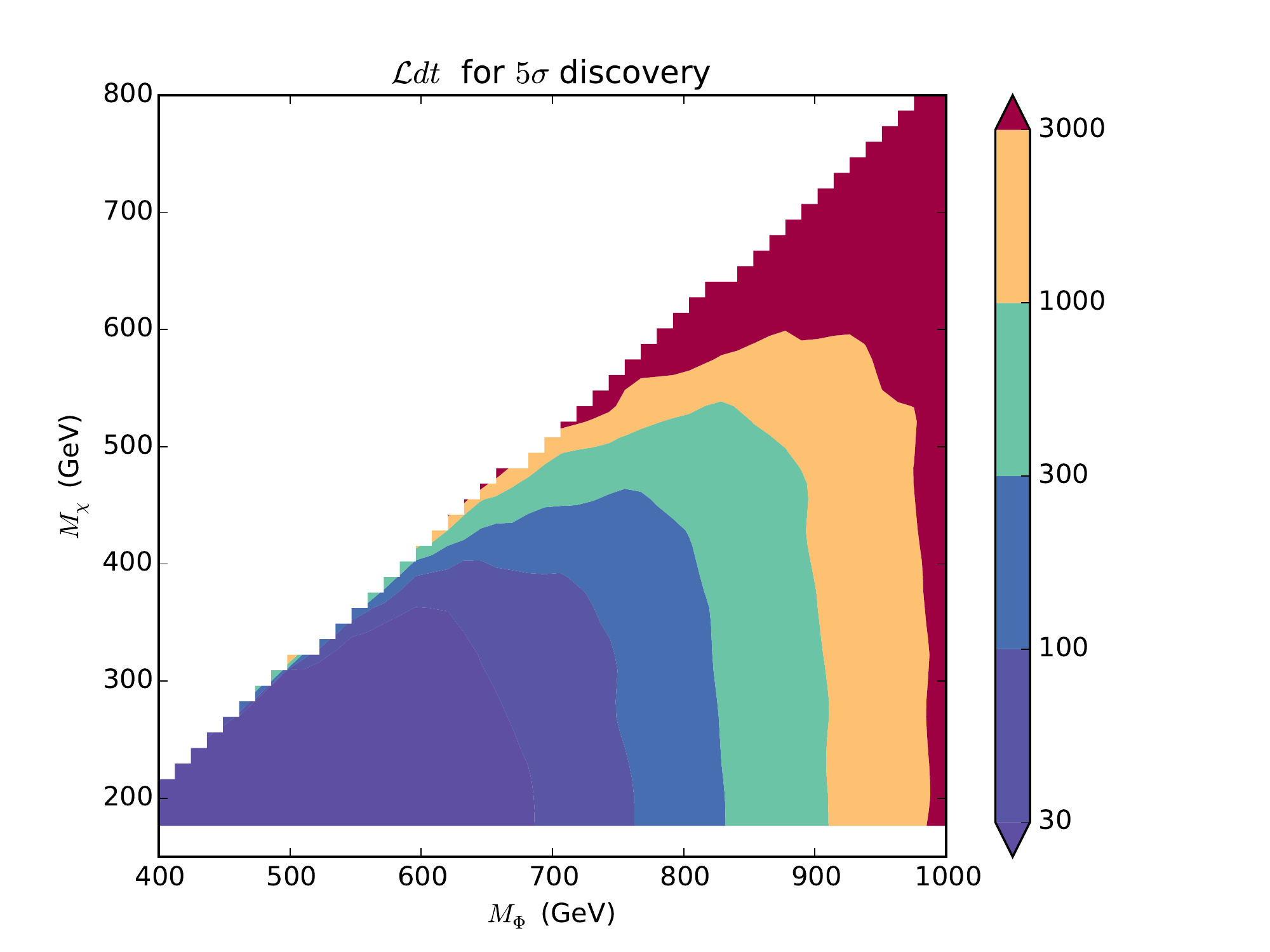} 
\includegraphics[width=0.49\textwidth]{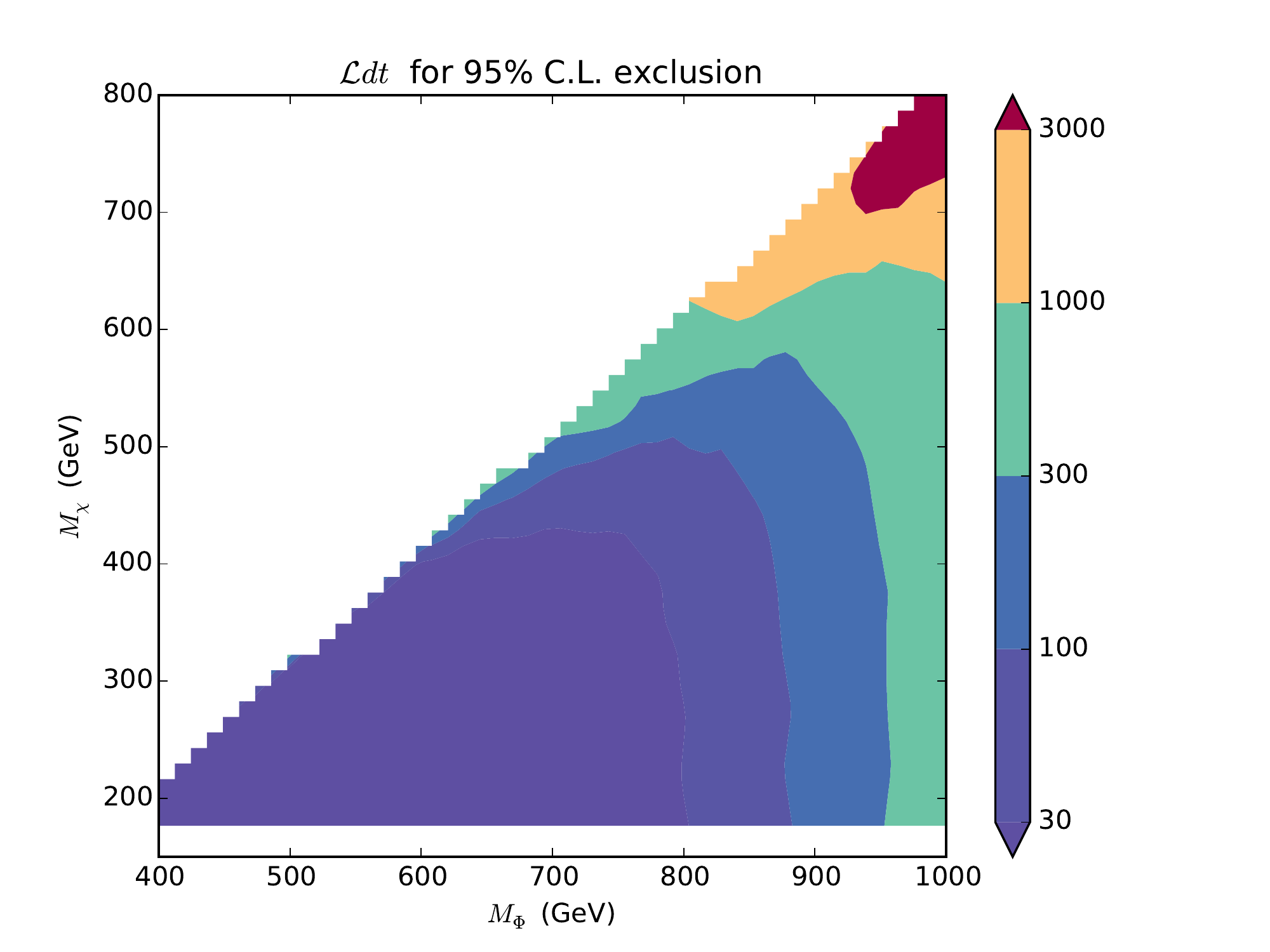} \\
\includegraphics[width=0.49\textwidth]{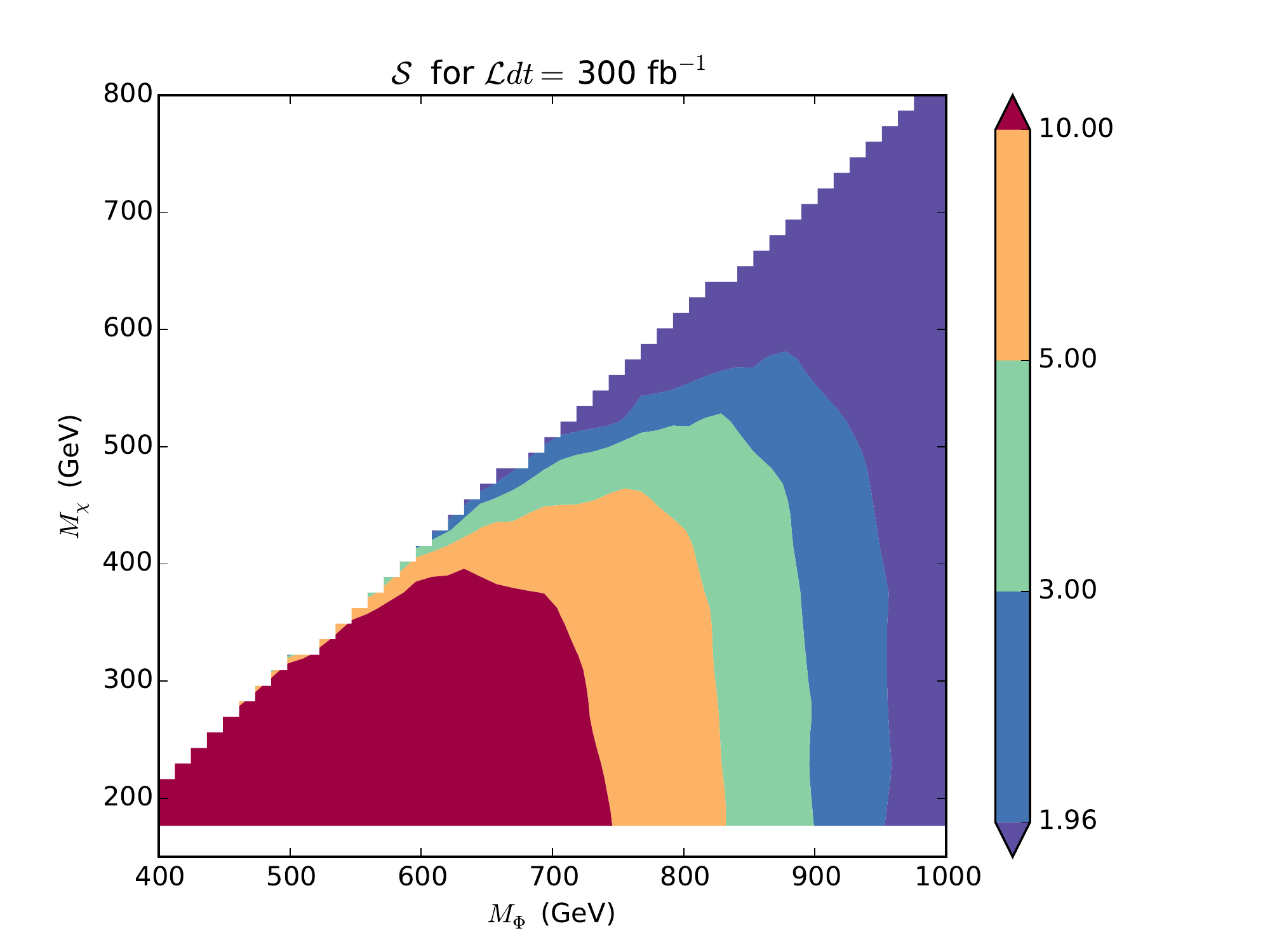} 
\caption{LHC reach in the $M_\phi$ - $M_\chi$ mass plane after optimizing cuts in the $M_T(b\ell,\met)$ vs. $\Delta \phi_{\ell, \met} $ plane.  Luminosity required for (a) $5\sigma$ discovery, (b) 95\% C.L. exclusion, and (c) expected statistical significance for $\int {\cal L} dt = 300 $ fb$^{-1}$.}
\label{fig:LHC}
\end{center}
\end{figure}

In practice, for the cut-based analysis, we optimize the statistical significance over the observables $M_T(b\ell,\met)$ and $\Delta \phi_{\ell,~\met} $, which offer good discrimination.   We define the level of statistical significance, ${\cal S}$, according to~\cite{Bartsch:2005xxa}
\be
{\cal S} = 2 \left(\sqrt{S+B}-\sqrt{B}\right),
\ee
where $S$ and $B$ are the number of signal in background events surviving cuts.  The  expected significance for $\int {\cal L} dt =  300 $ fb$^{-1}$ is presented in Fig.~\ref{fig:LHC} as are the luminosity required for 95\% C.L. exclusion and 5$\sigma$ discovery.

We extend our analysis to include multiple variables simultaneously.  This allows one to in essence blend cuts together rather than perform a hard cut on a kinematic distribution.  We form a discriminant based on a set of observables which include: ${\cal O}=\left\{ \met, M_T(b\ell,\met), \Delta \phi_{\ell~\met},M_T(\phi_h), M_T(\phi_\ell) \right\}$
\be
{\cal D}={S({\cal O})\over S({\cal O}) + A~ B({\cal O})}
\label{eqn:discrim}
\ee
where $S({\cal O})$ and $B({\cal O})$ are the normalized differential cross sections in the observable space ${\cal O}$.  These differential cross sections are estimated via event generation.  The discriminator is evaluated for an event sample, yielding a value close to 1 for signal-like events and close to 0 for background-like events.  For the particular choice of $A=N_B / N_S$, the discriminant gives the probability of an event being signal~\cite{Barlow:1986ek}.  A cut may be placed on the value of ${\cal D}$, thereby selecting a relatively high signal event sample.  Such a multivariate discriminator can offer similar sensitivity that the matrix-element, or neural network methods allow~\cite{Abazov:2008kt}.

\begin{figure}[htbp]
\begin{center}
\includegraphics[width=0.49\textwidth]{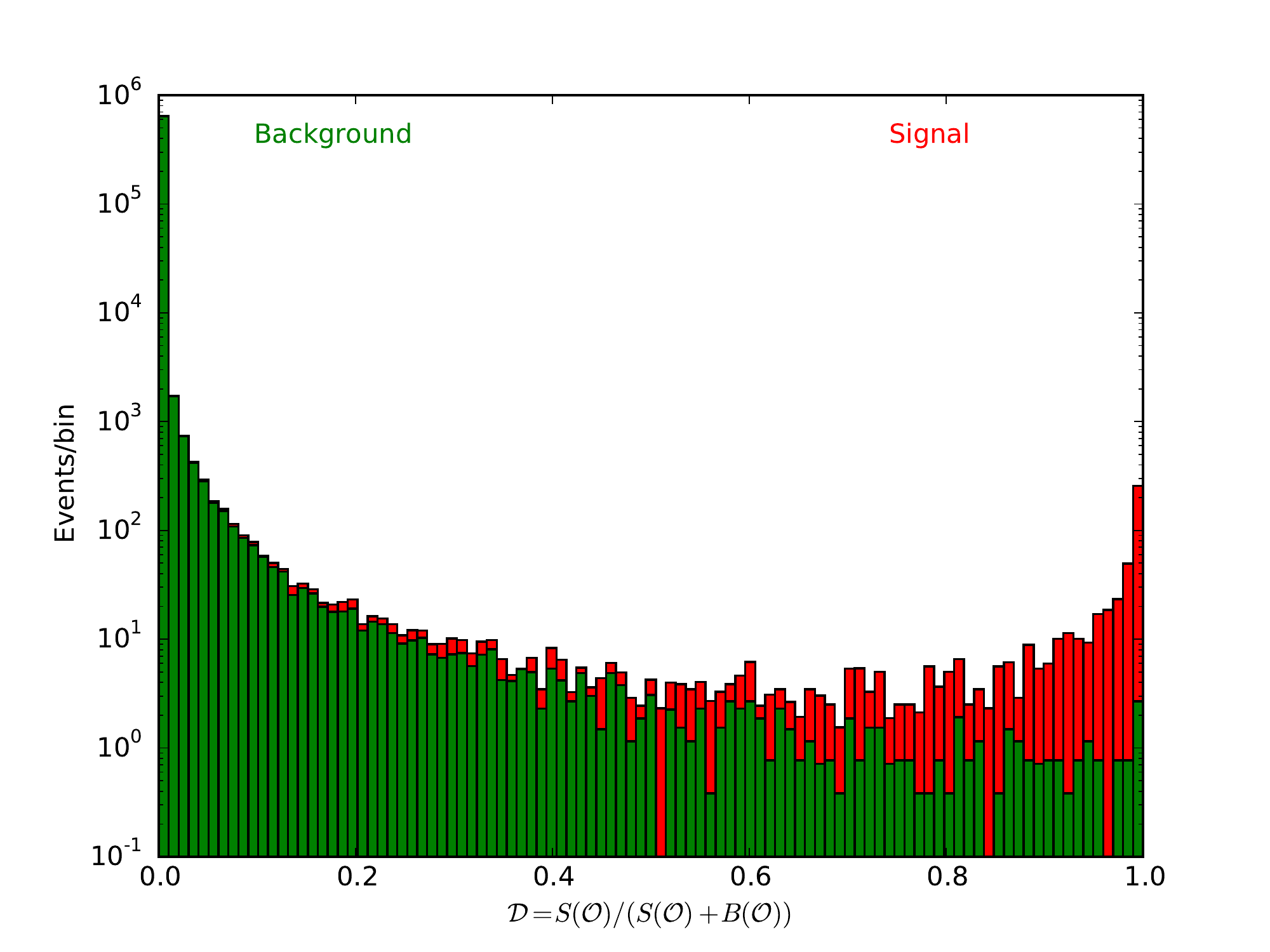}
\includegraphics[width=0.49\textwidth]{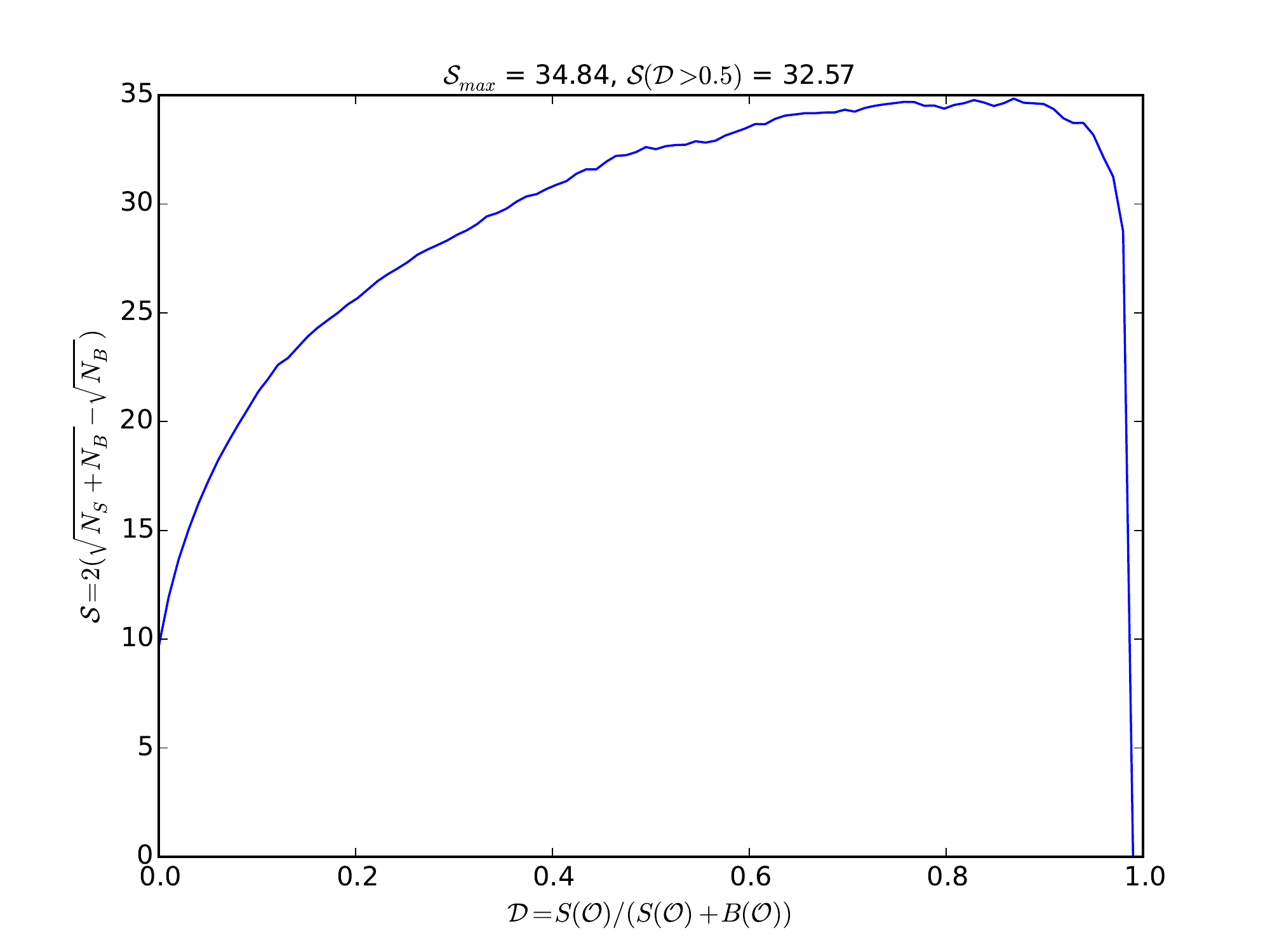}\\
\caption{Signal and background distribution over the MVA discriminant for a representative point in parameter space with $M_\phi=625$ GeV and $M_\chi = 225$ GeV with $\int{\cal L} dt =$ 300 fb$^{-1}$ using the observables $ \met, M_T(b\ell,\met)$, and $ \Delta \phi_{\ell~\met}$.  }
\label{fig:mva}
\end{center}
\end{figure}
In practice, we apply a simplified version of the discriminant in which we ignore the correlations among the variables.  With limited statistics, this allows a more efficient construction of the discriminator, defined as
\be
{\cal D}= {S\{{\cal O}_i\}\over S\{{\cal O}_i\} + B\{{\cal O}_i\}},
\ee
where $\{{\cal O}_i\}$ is the combinatorial subset of observables, ${\cal O}$ that  go into the multivariate discriminant. In the MVA results that follow, further optimization may be done by including the correlations between observables, but we adopt this uncorrelated approach for simplicity.  We maximize the significance, ${\cal S}$, by varying the cut on the discriminator, ${\cal D}_{\rm cut}$, this minimizes the choice of $A$ in Eq.~\ref{eqn:discrim}. We show, in Fig.~\ref{fig:mva}, the signal and background distribution in the discriminant for a test point in parameter space to illustrate how well the discriminant separates the two components.

\begin{figure}[htbp]
\begin{center}
\includegraphics[width=0.49\textwidth]{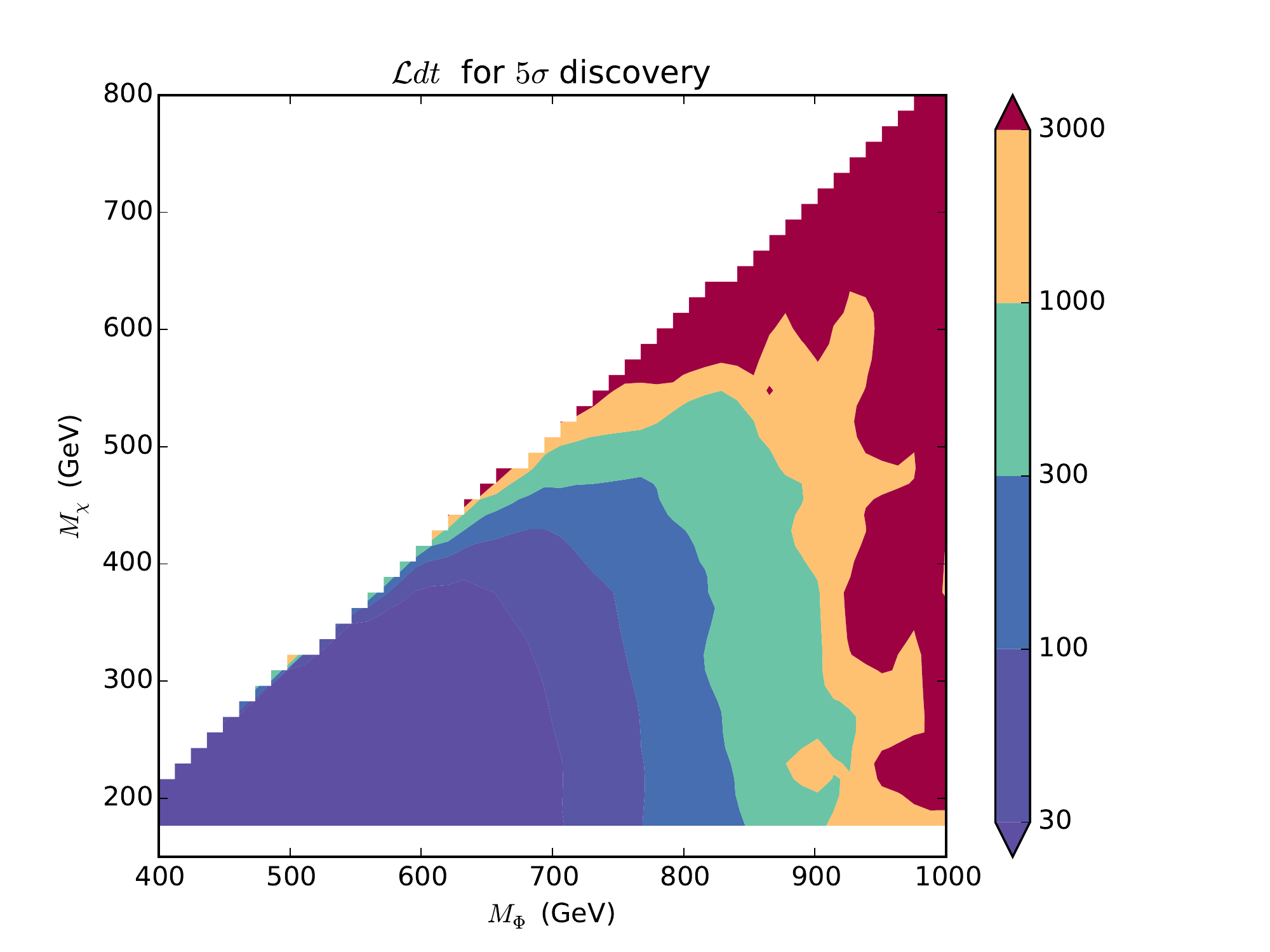} 
\includegraphics[width=0.49\textwidth]{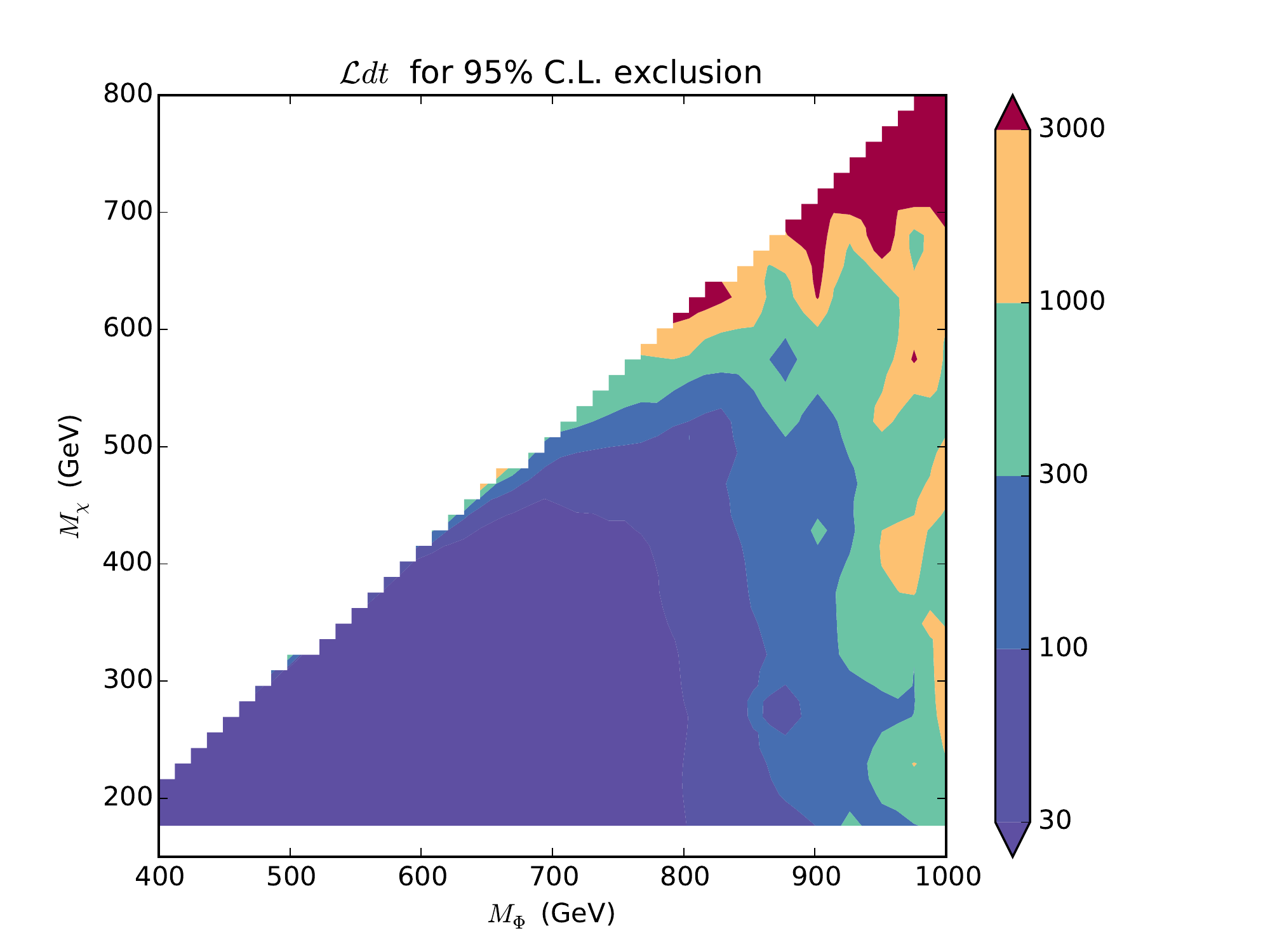} \\
\includegraphics[width=0.49\textwidth]{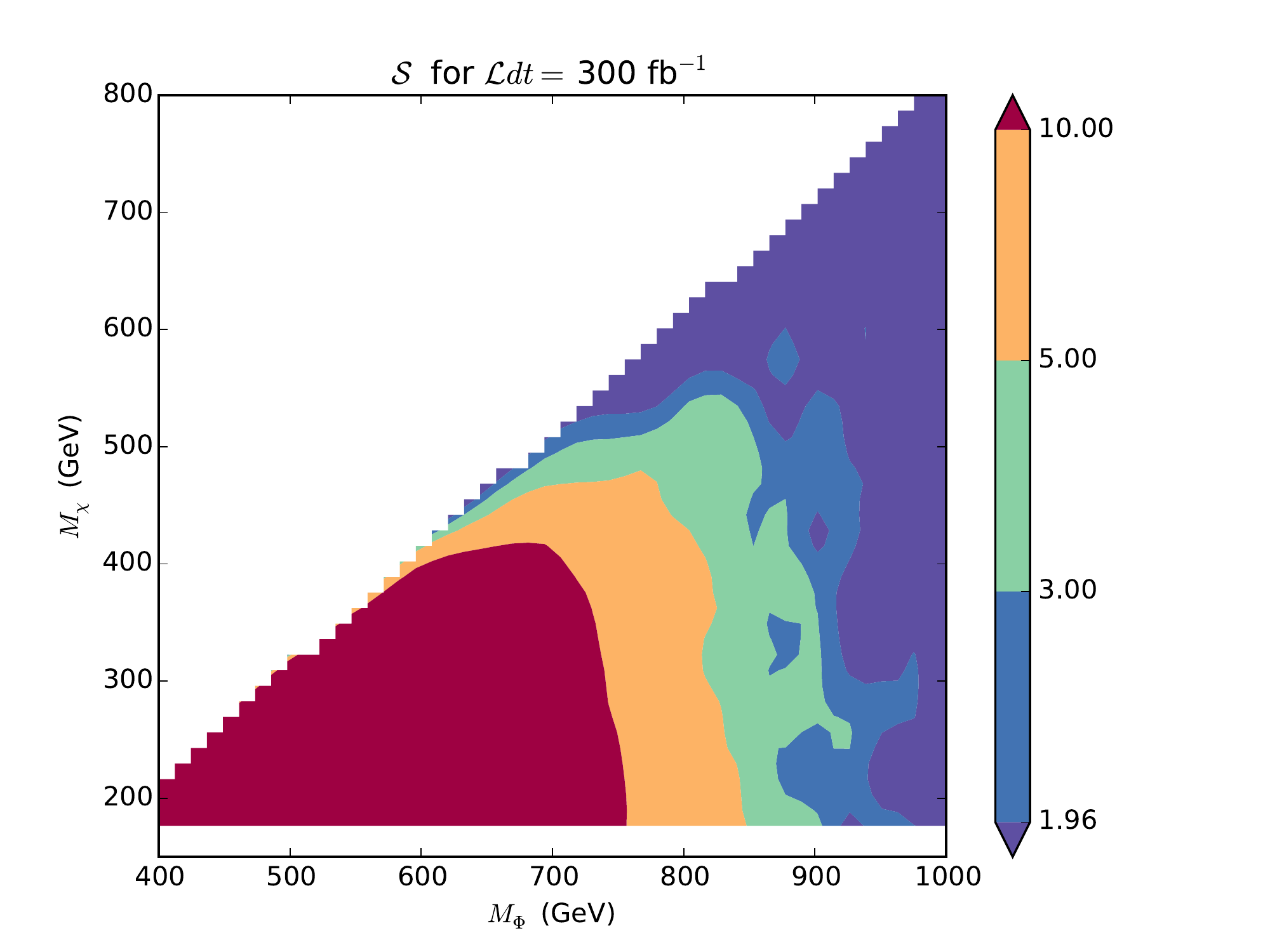} 
\caption{LHC reach in the $M_\phi$ - $M_\chi$ mass plane with the MVA.  Luminosity required for (a) $5\sigma$ discovery, (b) 95\% C.L. exclusion, and (c) expected statistical significance for $\int {\cal L} dt = 300 $ fb$^{-1}$.}
\label{fig:LHCMVA}
\end{center}
\end{figure}
The MVA results in the $M_\phi$ - $M_\chi$ plane are presented in~\ref{fig:LHCMVA}.  We note that the uncertainties are rather large for the higher luminosity (lower significance) contours.  This is due to the large number of events required to smooth the differential cross sections in Eq.~(\ref{eqn:discrim}).  In practice, this is difficult to achieve for the $t\bar t$ background sample for a large luminosity.  We generate $1.2\times10^7$ background events that pass the $\met$ and other threshold cuts.  However, even with these fluctuations, we can conclude that when comparing with the optimized cut based approach there are minor improvements in the significance as indicated when comparing the $5\sigma$ contour.  Moreover, the similarities between the MVA and optimized cut-based illustrate the choice of the $M_T(b\ell,\met)$ and $ \Delta \phi_{\ell~\met}$ observables in achieving a high significance.

\section{Indirect Detection from Annihilation into Gamma Rays}
\label{sec:indirect-det}

In this section, we consider the prospects of detecting TPDM indirectly from its annihilation into gamma rays.  Gamma rays from dark matter annihilations can come from several different processes.  Annihilation into charged SM particles can result in gamma rays from final-state radiation (FSR).  Also possible are gamma rays from annihilation into SM final states that subsequently hadronize into neutral pions which then decay into pairs of photons.  The combined spectra from FSR and $\pi^0$ decay result in a continuous and rather featureless spectrum.     Finally, annihilations of DM particles {\it directly} into $\gamma + X$ final states can occur at the loop level.  Because WIMPs are non-relativistic, the gamma rays produced from these annihilations manifest themselves as ``lines'' in the gamma ray spectrum.  Naively, one would expect that loop-level processes would be highly suppressed compared to continuum emission.  However, in models where DM has largish couplings to SM states and the particles in the loops have comparable masses to DM (so that threshold enhancements occur), the line emission can be comparable to (or even dominate over) the continuum.  These line emissions would provide a ``smoking gun'' for DM annihilation since it is believed that astrophysical processes are incapable of producing such features.  Unfortunately, the LUX constraints tell us that TPDM must be relatively heavy such that the detection of a signal from gamma rays with the Fermi LAT seems unlikely.  However, there are several gamma-ray telescopes in development including GAMMA-400 \cite{Galper:2012ji,Galper:2012fp}, CTA \cite{Consortium:2010bc} or HESS-II \cite{Becherini:2012yq} which will be able to probe these higher energies at very good experimental resolutions (see Ref.~\cite{Bergstrom:2012vd} and references therein for a full review).

Since TPDM is a Dirac fermion, its loop annihilations will result in final states consisting of ({\it i}) two photons, ({\it ii}) one photon plus one $Z$ boson as well as ({\it iii}) one photon plus one Higgs boson.  We will consider each of these final states after discussing the computation of the continuum.

\subsection{Gamma-ray Continuum from WIMP Annihilation}
\label{subsec:continuum}

To compute the gamma-ray continuum, we have implemented the model in micrOMEGAs \cite{Belanger:2013oya}.  In the TPDM model with non-degenerate $\chi$ and $\phi$ masses, pairs of WIMPs annihilate exclusively into pairs of top quarks.  The continuum spectra can, in general, be split into two regions.  At low values of $x$ (where $x = E_\gamma/M_\chi$), the main contribution comes from the decays of $\pi^0$s from the hadronization of strongly-interacting decay products.  For larger values of $x$, the main contribution comes from final-state radiation (FSR) which can be well-approximated by \cite{Bringmann:2007nk}:
\bea
\frac{dN}{dx} \approx \frac{\alpha_{EM} Q_t^2}{\pi} \left[ \frac{1 + (1 - x)^2}{x} \right] \log \left( \frac{s (1 - x)}{m_t^2} \right) \,,
\label{eq:FSR}
\eea
where $s \simeq 4 M_\chi^2$.  

\begin{figure}[t]
\begin{center}
\includegraphics[scale=0.5]{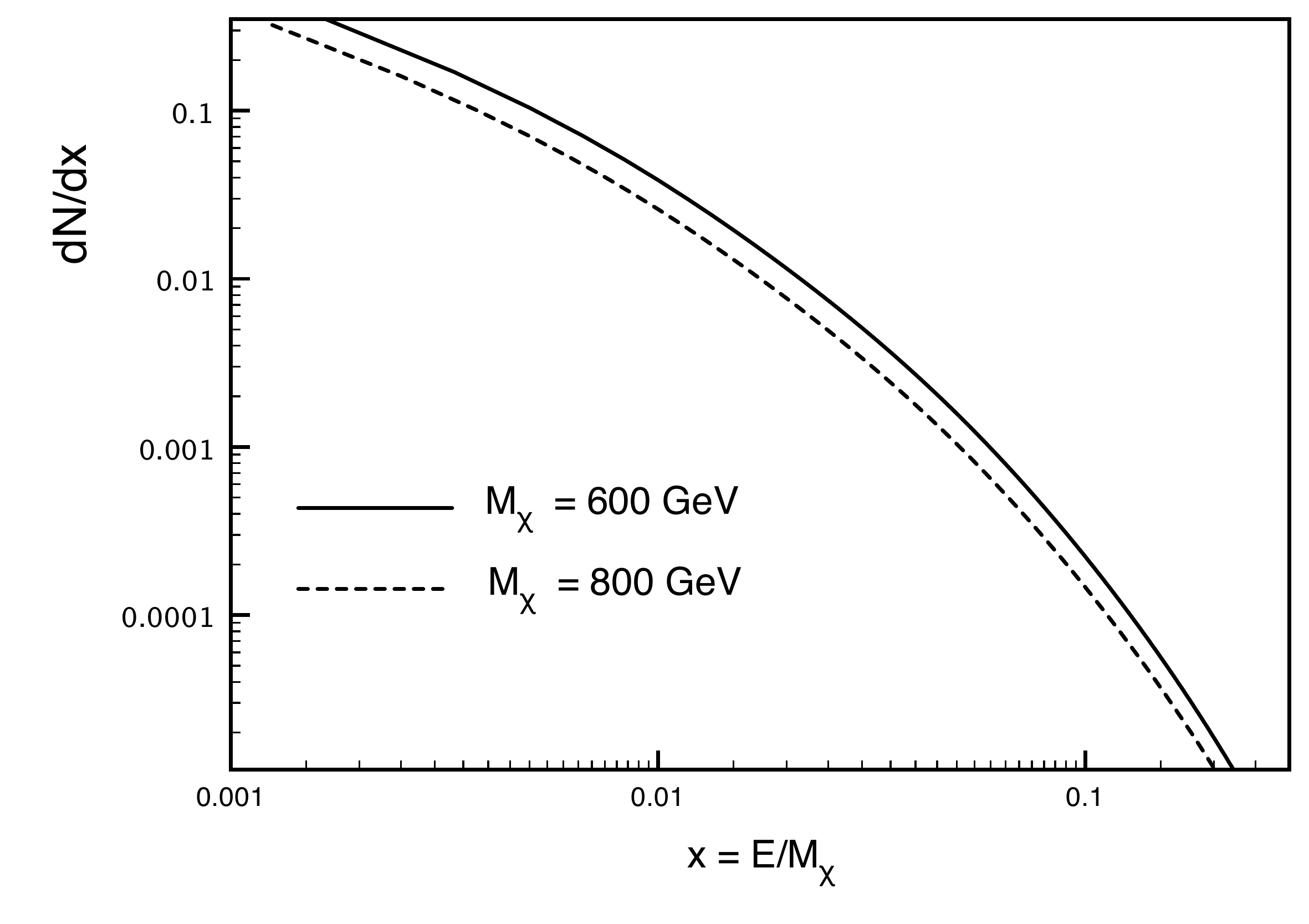} 
\caption{Continuum photon spectra for two WIMP masses. }
\label{fig:dNdx}
\end{center}
\end{figure}

In Fig.~\ref{fig:dNdx} we show a couple of (normalized) continuum photon spectra in the TPDM model for several choices of the WIMP mass.  From these spectra, we note they are quite similar in shape with a predominantly soft component indicative of the $\gamma$ originating from a showering and hadronizing top quark.  This softness is important for the prominence of the $\gamma$ lines that we compute in the following sections.
 
\subsection{Annihilation into Di-photon Final State}
\label{subsec:AACrossSec}

\begin{figure}[t]
\begin{center}
\includegraphics[scale=0.2]{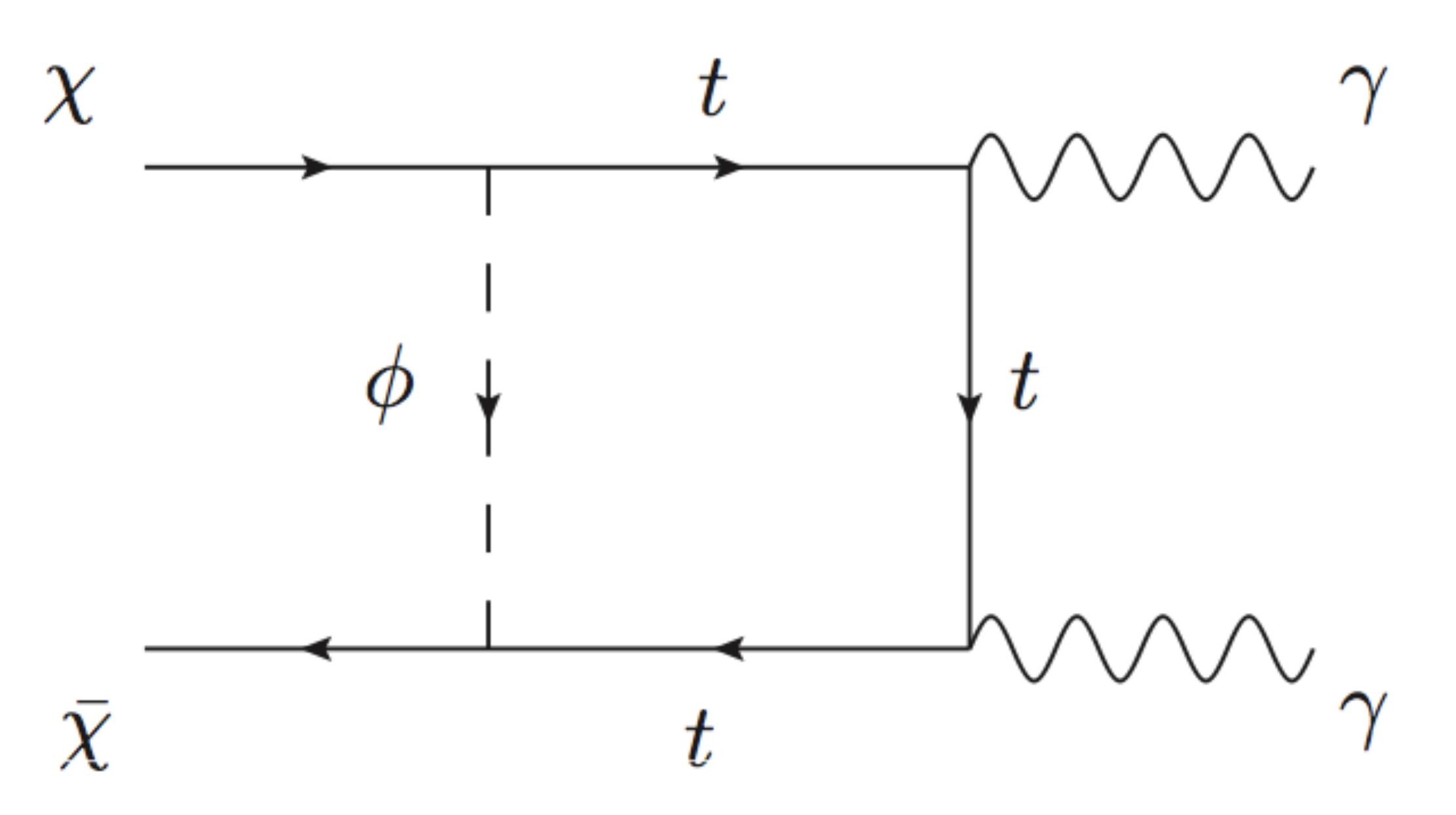} 
\includegraphics[scale=0.2]{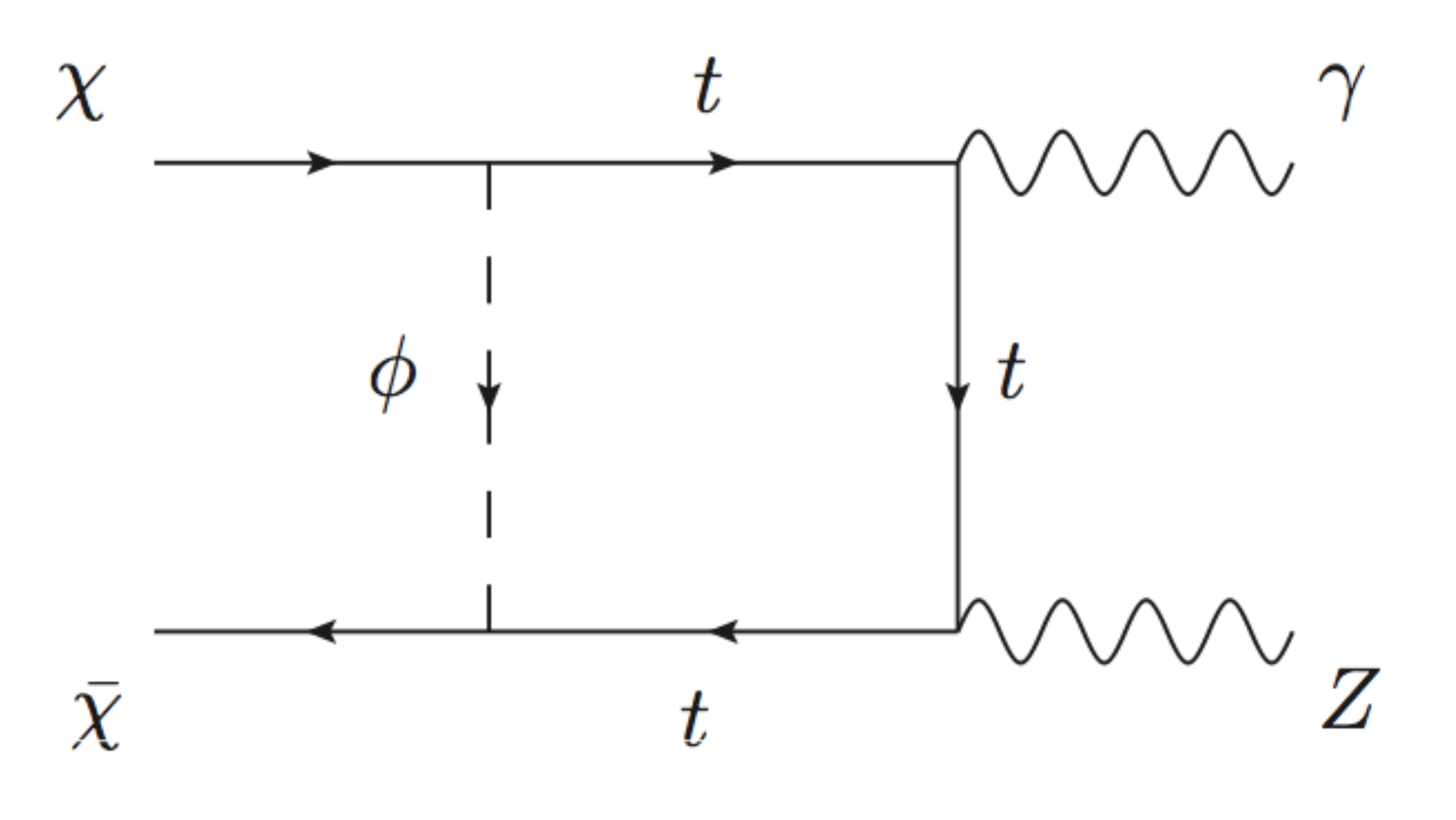} 
\includegraphics[scale=0.2]{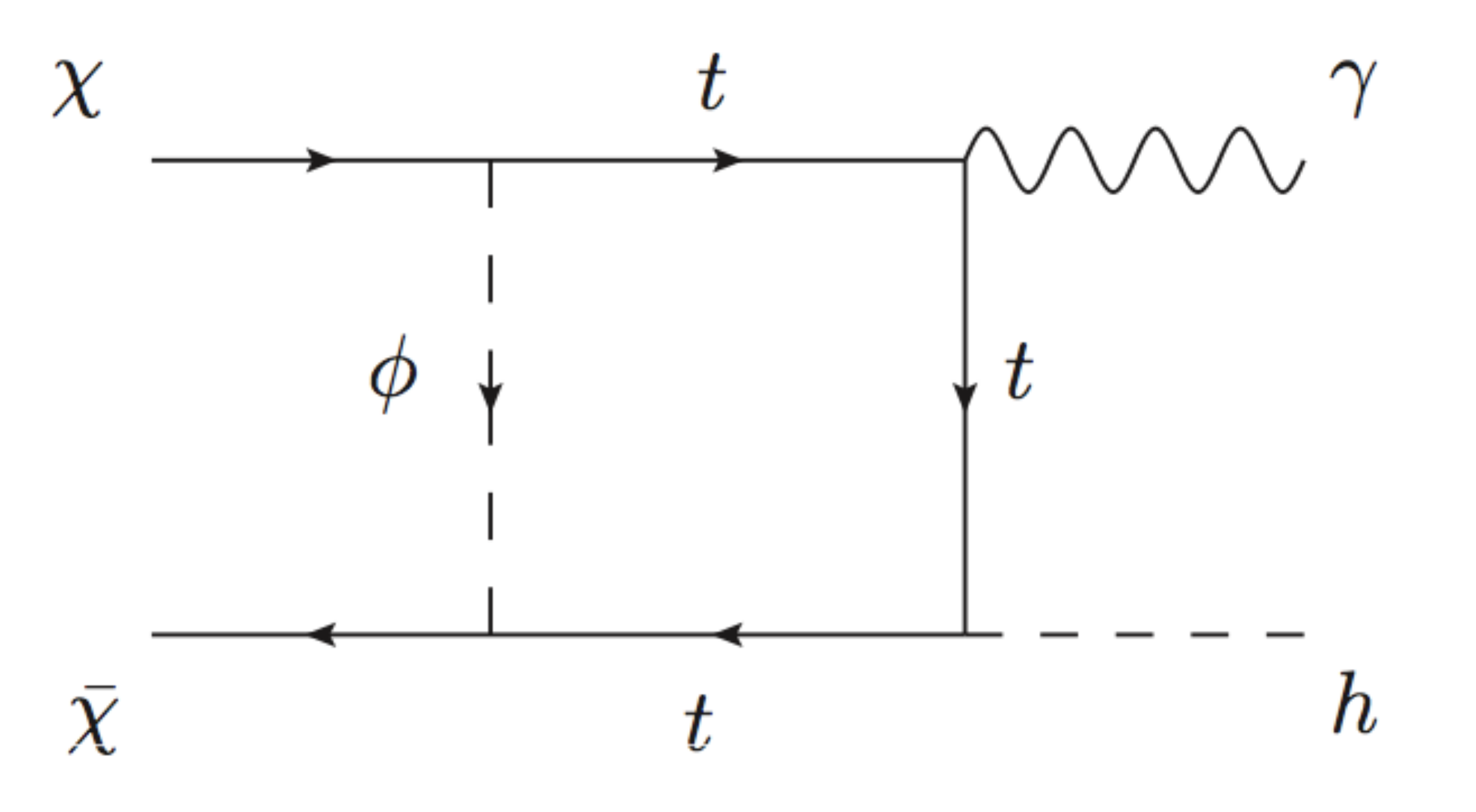} 
\caption{Representative Feynman diagrams which contribute to the processes $\chi \bar{\chi} \to \gamma \gamma$, $\chi \bar{\chi} \to \gamma Z$ and $\chi \bar{\chi} \to \gamma h$ respectively.}
\label{fig:AXdiagrams}
\end{center}
\end{figure}

The annihilation of TPDM into a pair of gamma rays proceeds via loops of top quarks and $\phi$ particles as depicted in the first Feynman diagram of Fig.~\ref{fig:AXdiagrams}.  The amplitude for the process $\chi(p_1) + \bar{\chi}(p_2) \to \gamma^\mu(p_A) + \gamma^\nu(p_B)$ can be written as:
\bea
{\cal M}^{\gamma\gamma} = \epsilon^{\mu *}(p_A) \epsilon^{\nu *}(p_B) {\cal M}^{\gamma\gamma}_{\mu\nu} \,,
\eea
where $\epsilon^\mu$ and $\epsilon^\nu$ are the polarization vectors for the two final-state photons.  The sub-amplitude ${\cal M}_{\mu\nu}$ can be expanded as a linear combination of tensor structures made up of the metric tensor as well as the external momenta.  Since the WIMPs are non-relativistic, we take the incoming momenta to be identical ($p_1 \simeq p_2 \equiv p = (M_\chi, {\bf 0})$) which greatly reduces the number of possible tensor structures in the amplitude.  Additionally, taking into account the transversality of the photons (i.e., $\epsilon(q) \cdot q = 0$) we are able to eliminate even more terms and are left with:
\bea
{\cal M}^{\gamma\gamma}_{\mu\nu} &=& \alpha_{DM} \alpha_{EM} Q_t^2 N_c \, \bar{v}(p) \, A^{\gamma\gamma}_{\mu\nu} \, u(p) \,,
\label{eq:Mgamgam}
\eea
where $\alpha_{DM} \equiv \frac{g_{DM}^2}{4\pi}$, $\alpha_{EM} \equiv \frac{e^2}{4\pi}$ and:
\bea
A^{\gamma\gamma}_{\mu\nu} &=& C_1 g_{\mu\nu} + C_2 \gamma_5 g_{\mu\nu} + C_3 \not{p}_A g_{\mu\nu} + C_4  \not{p}_A \gamma_5 g_{\mu\nu} \nonumber\\
&+& C_5 \gamma_\mu \gamma_\nu + C_6 \gamma_\mu \gamma_\nu \gamma_5 + C_7 \gamma_\mu \gamma_\nu \not{p}_A + C_8 \gamma_\mu \gamma_\nu \not{p}_A \gamma_5 \,.
\label{eq:Agamgam}
\eea

For the diagrams of interest, the coefficients $C_i$ are functions of scalar integrals as well as tensor coefficients (up to rank-three).  Following the usual Passarino-Veltman (PV) algorithm \cite{Passarino:1978jh}, these tensor coefficients can be reduced to functions of scalar integrals.  However, in the case where two of the incoming momenta are identical (the situation that arises in WIMP annihilation), the PV scheme breaks down and one is forced to augment the PV approach.  We choose to use the algebraic reduction scheme discussed in Ref.~\cite{Stuart:1987tt}.  For a more detailed discussion of the implementation of this scheme, see Ref.~\cite{Bertone:2009cb}.

The annihilation cross section is given by:
\bea
\langle \sigma v \rangle_{\gamma\gamma} = \frac{1}{64 \pi M_\chi^2} \left| {\cal M}^{\gamma\gamma} \right|^2 \,,
\eea
where we have included a factor of 1/2 in the cross section to account for the identical particles in the final state.

\subsection{Annihilation into a Photon Plus $Z$ Final State}
\label{subsec:ZACrossSec}

In the case of $\chi(p_1) + \bar{\chi}(p_2) \to \gamma^\mu(p_A) + Z^\nu(p_Z)$ which is depicted in the middle panel of Fig.~\ref{fig:AXdiagrams}, we can again write the amplitude as:
\bea
{\cal M}^{\gamma Z} = \epsilon^{\mu *}(p_A) \epsilon^{\nu *}(p_Z) {\cal M}^{\gamma Z}_{\mu\nu} \,,
\eea
with:
\bea
{\cal M}^{\gamma Z}_{\mu\nu} &=& \frac{\alpha_{DM} \alpha_{EM} Q_t}{4 s_w c_w} N_c \, \bar{v}(p) \, A^{\gamma Z}_{\mu\nu} \, u(p) \,.
\label{eq:Mgamgam}
\eea
However, due to the additional, longitudinal polarization of the $Z$ boson, we have additional terms to account for in the sub-amplitude:
\bea
A^{\gamma Z}_{\mu\nu} &=& C^{\gamma Z}_1 g_{\mu\nu} + C^{\gamma Z}_2 \gamma_5 g_{\mu\nu} + C^{\gamma Z}_3 \not{p}_A g_{\mu\nu} + C^{\gamma Z}_4  \not{p}_A \gamma_5 g_{\mu\nu} \nonumber\\
&+& C^{\gamma Z}_5 \gamma_\mu \gamma_\nu + C^{\gamma Z}_6 \gamma_\mu \gamma_\nu \gamma_5 + C^{\gamma Z}_7 \gamma_\mu \gamma_\nu \not{p}_A + C^{\gamma Z}_8 \gamma_\mu \gamma_\nu \not{p}_A \gamma_5  \nonumber\\
&+& C^{\gamma Z}_9 \gamma_{\mu} p_\nu + C^{\gamma Z}_{10} \gamma_\mu \gamma_5 p_\nu + C^{\gamma Z}_{11} \gamma_{\mu} p_{A,\nu} + C^{\gamma Z}_{12}  \gamma_\mu \gamma_5 p_{A,\nu}  \nonumber\\
&+& C^{\gamma Z}_{13} \gamma_\mu \not{p}_A p_\nu + C^{\gamma Z}_{14} \gamma_\mu \not{p}_A \gamma_5 p_\nu + C^{\gamma Z}_{15} \gamma_\mu \not{p}_A p_{A,\nu} + C^{\gamma Z}_{16} \gamma_\mu \not{p}_A \gamma_5 p_{A,\nu} \,.
\label{eq:Agamgam}
\eea

The cross section for annihilation into a photon plus $Z$ boson final state is given by:
\bea
\langle \sigma v \rangle_{\gamma Z} = \frac{1}{32 \pi M_\chi^2} \left( 1 - \frac{m_Z^2}{4 M_\chi^2} \right) \left| {\cal M}^{\gamma Z} \right|^2 \,.
\eea

\subsection{Annihilation into a Photon Plus $h$ Final State}
\label{subsec:hACrossSec}

Finally, we consider the cross section for annihilation to a photon plus SM Higgs boson final state (as depicted in the far right Feynman diagram of Fig.~\ref{fig:AXdiagrams}).  As discussed earlier, this final state is possible because the WIMP we are considering is fermionic.  In addition, because the Higgs is radiated from a top quark line, the possibility for an enhanced cross section (compared to continuum and/or other gamma-ray lines) exists such that observation of a ``Higgs in space'' could be a reality~\cite{Jackson:2009kg,Jackson:2013pjq,Jackson:2013tca}.

The amplitude for the process $\chi(p_1) + \bar{\chi}(p_2) \to \gamma^\mu(p_A) + h(p_h)$ which is depicted in the far right diagram of Fig.~\ref{fig:AXdiagrams} can be written as:
\bea
{\cal M}^{\gamma h} = \epsilon^{\mu *}(p_A) {\cal M}^{\gamma h}_{\mu} \,,
\eea
where the sub-amplitude ${\cal M}^{\gamma h}_{\mu}$ is:
\bea
{\cal M}^{\gamma h}_{\mu} = \alpha_{DM} \sqrt{4\pi \alpha_{EM}} Q_t \left( \frac{m_t}{v} \right) N_c  \, \bar{v}(p) \, A^{\gamma h}_{\mu} \, u(p) \,.
\eea
The tensor $A^{\gamma h}_\mu$ can now be expanded in terms of the external momenta.  Again, taking into account the non-relativistic nature of the WIMPs and the transversality of the final-state photon, the tensor simplifies to:
\bea
A^{\gamma h}_\mu =  C_1^{\gamma h} \gamma_\mu + C_2^{\gamma h} \gamma_\mu \gamma_5 + C_3^{\gamma h} \gamma_\mu \not{p}_A + C_4^{\gamma h} \gamma_\mu \not{p}_A \gamma_5 \,.
\eea

The cross section for annihilation into a photon plus a SM Higgs boson final state is given by:
\bea
\langle \sigma v \rangle_{\gamma h} = \frac{1}{32 \pi M_\chi^2} \left( 1 - \frac{m_h^2}{4 M_\chi^2} \right) \left| {\cal M}^{\gamma Z} \right|^2 \,,
\eea
where we take $m_h = 126$ GeV.  

\subsection{The Gamma-ray Spectrum from WIMP Annihilation}
\label{subsec:gamma-spectrum}

Finally, we assemble the above contributions and compute the total gamma-ray flux originating from a region centered on the center of the Milky Way galaxy.  The differential flux of gamma rays arising from dark matter annihilation observed in a direction making an angle $\psi$ with the direction of the galactic center (GC) is given by:
\bea
\frac{d \Phi_\gamma}{d\Omega dE_\gamma}(\psi, E_\gamma) = \frac{r_\odot \rho_\odot^2}{4 \pi M_\chi^2} \frac{dN_\gamma}{dE_\gamma} \int_{\rm l.o.s.} \frac{ds}{r_\odot} \left[ \frac{\rho[r(s,\psi)]}{\rho_\odot} \right]^2,
\label{eq:diffFlux}
\eea
where $\rho(\vec{x})$, $\rho_\odot = 0.3$ GeV/cm$^3$ and $r_\odot = 8.5$ kpc respectively denote the dark matter density at a generic location $\vec{x}$ with respect to the GC, its value at the solar system location and the distance of the Sun from the GC.  The quantity $dN_\gamma / dE_\gamma$ represents the sum over all annihilation channels $f$ with the corresponding cross section $\langle \sigma v \rangle_f$:
\bea
\frac{dN_\gamma}{dE_\gamma} = \sum_f \langle \sigma v \rangle_f \frac{dN^f_\gamma}{dE_\gamma} \,,
\eea
where $\frac{dN^f_\gamma}{dE_\gamma}$ represents the normalized photon spectrum per annihilation event.  Note that, in the case of $\gamma Z$ and $\gamma h$, the photon spectrum will broaden from monochromatic emission due to the decay widths $\Gamma_X$ where $X = Z$ or a Higgs boson.  The exact expression is given by:
\bea
\frac{dN_\gamma^X}{dE} = \frac{4 M_\chi m_X \Gamma_X}{f_1 f_2} \,,
\eea
where:
\bea
f_1 &=& \tan^{-1} \left( \frac{m_X}{M_\chi} \right) + \tan^{-1} \left( \frac{4 M_\chi^2 - m_X^2}{m_X \Gamma_X} \right) \,, \\
f_2 &=& \left( 4 M_\chi^2 - 4 M_\chi E_\gamma - m_X^2 \right)^2 + \Gamma_X^2 m_X^2 \,.
\eea

In addition to accounting for the finite width of the $\gamma Z$ and $\gamma h$ lines, we also account for the finite resolution of the detector by convolving the ``raw'' photon flux from Eq.~(\ref{eq:diffFlux}) with a Gaussian kernel $G(E, E_0)$:
\bea
G(E, E_0) = \frac{1}{\sqrt{2\pi} E_0 \sigma} \exp \left[ - \frac{(E - E_0)^2}{2 \sigma^2 E_0^2} \right] \,,
\eea
where $\sigma$ is related to the detector's relative energy resolution $\xi$ by $\sigma = \xi/2.3$.

\begin{figure}[t]
\begin{center}
\includegraphics[width=0.75\textwidth]{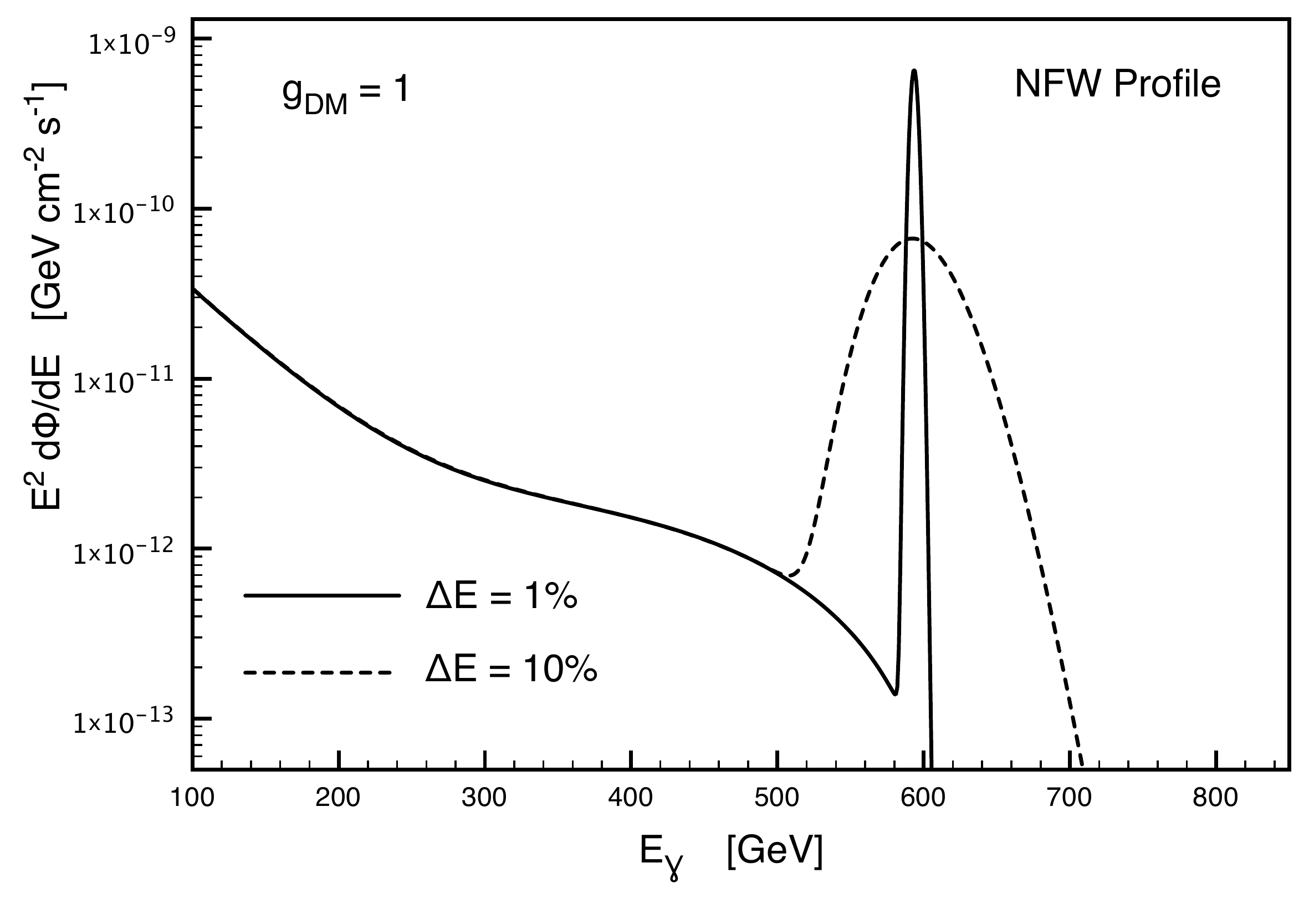} 
\caption{Gamma-ray spectra for $M_\chi = 600$ using the NFW profile for two experimental energy resolutions of $\Delta E = 1\%$ and $\Delta E = 10\%$.}
\label{fig:gamma-flux-mc600-DelEdep}
\end{center}
\end{figure}

In Fig.~\ref{fig:gamma-flux-mc600-DelEdep}, we plot the gamma-ray spectrum for a particular point in parameter space which results in the correct relic abundance ($g_{DM} = 1$, $M_\chi = 600$ GeV and $M_\phi = 1150$ GeV).  We plot the spectrum assuming two different experimental resolutions,  $\Delta E = 1\%$ (solid) and $\Delta E = 10\%$ (dashed), and using the Navarro-Frenk-White (NFW) DM density profile \cite{Navarro:1995iw}.  The first thing we notice is that, due to resolution detector effects, the three lines get smeared into one large bump.  However, the prominence of the bump over the background is quite impressive and not typical of any other model.  The main effect responsible for this is the suppression of the continuum as discussed above.


\section{Conclusions}
\label{sec:conclusion}

In this paper, we have considered a simplified model of dark matter where the WIMP is a Dirac fermion which couples exclusively to the SM right-handed top quark and a new scalar field, $\phi$, via a Yukawa-like interaction.  In order for this new interaction to preserve the SM gauge-invariance, $\phi$ must carry the SM quantum numbers of the right-handed top quark (i.e., it has both electroweak and strong interactions).  We call this type of dark matter ``top portal dark matter'' or TPDM.

We were motivated to study TPDM because of the seemingly coincidental similarity between ({\it i}) the scale of EWSB which is $\sim{\cal O}(100~{\rm GeV})$ as evidenced by the recent discovery of a Higgs boson consistent with that predicted by the SM and ({\it ii}) the scale of WIMP dark matter which is believed to be $\sim{\cal O}(100~{\rm GeV}-10~{\rm TeV})$.  If there is a connection between EWSB and DM, it could be possible that DM will have enhanced couplings to the heaviest objects in the SM such as the top quark much like the SM Higgs boson.

One of the attractive features of studying simplified models such as the one we considered is the small number of free parameters; for TPDM, there are only three (the coupling for the Yukawa-like interaction, the mass of the WIMP and the mass of the scalar).  One way to constrain the parameter space of TPDM is through the measured relic abundance of DM.  In Section~\ref{sec:relic-density}, we computed the current density of TPDM assuming it is a thermal relic.  Comparing our predictions with data, we showed that TPDM can explain the relic abundance for a wide range of parameters.  Furthermore, requiring TPDM to saturate the relic density allowed us to determine the coupling given the WIMP and scalar mass.

Next, we considered the possibility of detecting TPDM at direct detection experiments.  The null results from current direct detection experiments such as the LUX 10000 kg da limit the mass ranges available while simultaneously providing the observed WIMP relic abundance.  Generally, masses below $M_\chi < 450$ GeV and $M_\phi < 750$ GeV are ruled out, assuming no experimental uncertainty in the location of the NR band. Including the $1\sigma$ uncertainty of this band, these limits are 400 GeV and 600 GeV, respectively.  

We also performed a full analysis of the possibilities of detecting TPDM at the LHC.  We found that the main production channel is through pair production of the colored scalars (through gluon annihilation) with each of the scalars subsequently decaying into a top quark and a WIMP.  This production rate is largely unaffected by the $\phi-\chi-t$ coupling.  The main SM background to this signal is the pair production of top quarks which can be quite significant at the energies considered here.  However, through both our cut-based and MVA analyses, we have found several discriminating physical observables which allow for a highly-significant discovery of TPDM at the LHC.

Finally, we considered the annihilation of TPDM into final states involving gamma rays.  We computed the expected flux of gamma rays originating from the galactic center.  This flux has several components: ({\it i}) a continuum from annihilations into SM final states which then radiate photons and/or hadronize/decay into photons and ({\it ii}) the direct (loop-level) annihilation into $\gamma + X$ final states which manifest themselves as ``lines'' in the gamma-ray spectrum.  Because the main annihilation mode of TPDM is into pairs of top quarks (and because the relic density constrains the WIMP mass to be not much more than the top mass), we found that the gamma-ray continuum from WIMP annihilations is highly-suppressed.  However, the suppressed continuum along with relatively large cross sections for direct annihilations into $\gamma + X$ final state, result in a total spectrum which exhibits a very striking and prominent bump which may be probed with next generation gamma ray telescopes.

\section{Acknowledgements}
We thank Yang Bai, Brian Batell, Geraldine Servant and Marco Taoso for helpful discussions.  G.S. is supported by the U. S. Department of Energy under the contract DE-FG-02-95ER40896.

\end{document}